\documentclass[a4paper,11pt, aastex631]{article}
\pdfoutput=1 

\usepackage{jcappub} 

\usepackage[]{fontenc}
\usepackage{amsmath}
\usepackage{amssymb}

\newcommand{\cii}{C~{\scriptsize II}}

\title{Dust in High-Redshift Galaxies: Reconciling UV Attenuation and IR Emission}

\author[a, b, 1]{Roy J. Zhao, \note{Corresponding author}}
\author[a]{Steven R. Furlanetto}

\affiliation[a]{Department of Physics \& Astronomy, University of California, Los Angeles, \\475 Portola Plaza, Los Angeles, CA 90095-1547, USA}
\affiliation[b]{Department of Physics, The University of Chicago, \\5720 S Ellis Ave, Chicago, IL 60637, USA}


\emailAdd{rzhaolx@ucla.edu}

\abstract{Dust is a key component of galaxies, but its properties during the earliest eras of structure formation remain elusive. Here we present a simple semi-analytic model of the dust distribution in galaxies at $z \gtrsim 5$. We calibrate the free parameters of this model to estimates of the UV attenuation (using the IRX-$\beta$ relation between infrared emission and the UV spectral slope) and to ALMA measurements of dust emission. We find that the observed dust emission requires that most of the dust expected in these galaxies is retained (assuming a similar yield to lower-redshift sources), but if the dust is spherically distributed, the modest attenuation requires that it be significantly more extended than the stars. Interestingly, the retention fraction is larger for less massive galaxies in our model. However, the required radius is a significant fraction of the host's virial radius and is larger than the estimated extent of dust emission from stacked high-$z$ galaxies. These can be reconciled if the dust is distributed anisotropically, with typical covering fractions of $\sim 0.2-0.7$ in bright galaxies and $\lesssim 0.1$ in fainter ones. }

\begin{document}
\maketitle
\flushbottom

\section{Introduction}
\label{sec:intro}
Dust is crucial for understanding galaxy evolution on a variety of levels. Not only is it a key component in chemical evolution, but it also absorbs a substantial fraction of UV starlight and re-emits it in the infrared. However, its role in galaxies during the Cosmic Dawn is mostly unknown, as only recently have the tools emerged to measure its properties. 

JWST has, of course, made crucial contributions by pushing the observational frontier to higher redshifts and fainter sources (e.g., \cite{Atek2023, Austin2023, Bouwens2023, Endsley2023, Tacchella2023, Topping2023a, Whitler2023}). Interestingly, these studies have found an excess of UV-luminous galaxy candidates at $z \sim 10$ compared to HST-calibrated models \cite{Mason2023, Mirocha2023, Ferrara2023}. The most popular explanation is burstiness in the star formation histories \cite{Mason2023}. However, dust may play a further role in this phenomenon \cite{Ferrara2023, Ziparo2023} if the highest-redshift sources are mostly dust-free. \cite{Mirocha2023} also argued that the burst solution also depends on dust: more efficient dust production is needed to match the JWST-observed UV colours while maintaining a star formation efficiency that fits the observed luminosity function. Others have noted that the dust may have a complex morphological relation to the star-forming regions \cite{Inami2022, Ziparo2023}. A thorough study of high-redshift dust properties is called for to answer these questions. 

Beyond the luminosity function, an understanding of dust is also crucial to interpreting galaxy properties during the Cosmic Dawn. Dust grains attenuate starlight, especially in the UV \cite{Calzetti2001, Weingartner2001, Schouws2022}. Therefore, it is critical for measuring the intrinsic SFR as well as other properties of the stellar population. However, most high-$z$ galaxies are only identified photometrically, and SED modelling is required to estimate the effects of dust. A common simple method is to use the UV spectral slope $\beta$ as a proxy for the dust distribution, which appears to be a good assumption for lower-redshift galaxies, although the details may evolve with redshift \cite{Meurer1999, Finkelstein2012, Casey2014, McLure2018, Reddy2018, Fudamoto2020, Bowler2023}. The fundamental motivation behind these empirical relations is that more dust (and hence a larger infrared luminosity) will redden the galaxy more (making the spectral slope flatter). 

The advent of ALMA has allowed new studies directly targeting dust emission in the far-infrared from high-$z$ galaxies \cite{Fudamoto2020, Khusanova2021, Sommovigo2021, Dayal2022, Ferrara2022, Inami2022, Fudamoto2022, Algera2023, Palla2023}. In particular, large surveys like ALPINE \cite{Bethermin2020, Faisst2020}, CRISTAL \cite{Mitsuhashi2023b}, SERENADE \cite{Mitsuhashi2023}, and REBELS \cite{Bouwens2022} have begun to place statistical constraints on the dust content of early galaxies. With the shared objective of studying dust and other ISM properties in high-$z$ galaxies, these surveys scan one or both of the brightest fine structure lines ([O III] at 88 $\mu$m and [\cii] at 158 $\mu$m) from UV or spectroscopically selected samples, which also allows a measurement of the nearby dust continuum. These surveys cover a broad range of redshifts (ALPINE and CRISTAL at $4.4\lesssim z\lesssim5.9$, SERENADE at $z\sim6$, and REBELS at $z\sim7$).

However, constructing a first-principles model of dust is a difficult task. It requires a prescription for dust formation in supernova ejecta \cite{Nozawa2003, Brooker2022}, AGB star winds \cite{Valiante2009}, ISM grain growth \cite{Mancini2015, Dayal2022}, and other processes. The dust must then be distributed through the interstellar and circumgalactic media through winds and turbulent transport, leaving a complex geometrical distribution \cite{Angle-Alcazar2017}. It then attenuates starlight in a manner that depends on the composition and size distribution of the dust grains \cite{Calzetti2001, Draine2003}. Recent work has brought a variety of methods to bear on this problem. Numerical simulations \cite{Narayanan2018, Vogelsberger2020, Mushtaq2023, Narayanan2023} can track the evolution of dust content in detail but are still dependent on assumptions about feedback and dust creation and destruction. Analytic models offer more flexibility but are limited in the number of physical mechanisms they can include \cite{dacunha2013, Imara2018, Inoue2020}. Many of these models use parameterized semi-analytic approaches to account for these challenges \cite{Sommovigo2021, Dayal2022, Ferrara2022, Mauerhofer2023, Palla2023}. Other models focus on studying the chemical composition of dust either through parametric approaches \cite{Calura2008, Mattsson2022} or in a cosmological context \cite{McKinnon2017, Hou2019, Parente2022}.

In this paper, we use the new JWST and ALMA observations to understand dust through a simplified model of galaxy formation. We use the ``minimalist'' model of high-$z$ galaxy evolution from \cite{Furlanetto2017} as our baseline and incorporate a new treatment of dust production and morphology. Although simple, this model reproduces properties of the galaxy population at $z<10$ well and is qualitatively similar to several other semi-analytic models \cite{Trenti2010, Dayal2014, Mason2015}. We then use both UV and far-IR measurements to constrain the properties of the model. 

The paper is organized as follows. In section \ref{Methods}, we introduce our galaxy formation and dust models, building on \cite{Furlanetto2017}; in section \ref{results}, we compare to UV dust measurements; in section \ref{ALMA}, we appeal to the ALMA far-IR flux observation to further constrain the model; in section \ref{further}, we explore the implications of our model; and in section \ref{conclusions} we conclude.

The numerical calculations in this work assume a flat $\Lambda$CDM cosmology with $\Omega_m = 0.3111$, $\Omega_\Lambda = 0.6889$, $\Omega_b = 0.0489$, $h = 0.6766$, consistent with recent results of \cite{Planck2018}. The magnitudes in this paper are expressed in the AB system \cite{Oke1983}.

\section{Dust in High-redshift Galaxies}
\label{Methods}

This section introduces our dust prescription, which extends the star formation model of \cite{Furlanetto2017}. Since our objective is to constrain the dust mass and radius using multi-band observations, we extend their minimalist approach to include both dust evolution and its observable consequences. Rather than attempt to model dust from first principles, we will introduce scaling parameters for the dust mass and radius, to be calibrated by observations in later sections. The resultant model depends on the host dark matter halo mass $M_{\rm halo}$ and redshift $z$ of a galaxy, generating its (1) dust mass, (2) dust radius, (3) UV attenuation, and (4) dust temperature and far-IR continuum emission flux. 

\subsection{The Minimalist Galaxy Formation Model}
Here, we introduce our basis model: a minimalist semi-analytic star formation model by \cite{Furlanetto2017}. Calibrated to the HST-observed UVLF, this model predicts various properties of a galaxy by balancing its star formation and feedback, while assuming galaxies grow through matter accretion onto their host halos. The model parameters are rigorously fit to the pre-JWST UVLF measurements in $z=6-9$ through a Markov Chain Monte Carlo (MCMC) procedure by \cite{Hegde&Wyatt2024}. However, we must be cautious applying them to earlier eras.

The model of \cite{Furlanetto2017} uses a simple approximation to dark matter halo growth. We first define the halo mass function, by assuming that the comoving number density $n_{\rm h}$ of halos in the infinitesimal mass range $(M, M+dM)$ at a given redshift $z$ is
\begin{equation}
    n_{\rm h}(M, z) = f(\sigma)\frac{\bar{\rho}}{M} \frac{d\ln{(1/\sigma)}}{dM} ,
\end{equation}
where $\bar{\rho}$ is the comoving matter density of halos, $\sigma$ is the RMS fluctuation of the linear density field smoothed on a scale $M$, and $f(\sigma)$ is $\sigma$-weighted distribution of random-walk barrier-crossings for N-body simulations. Here, we take the $f(\sigma)$ parametrization from \cite{Trac2015}'s SCORCH hydrodynamical high-$z$ simulation results, where
\begin{equation}
    f(\sigma) = 0.15\left[1+\left(\frac{\sigma}{2.54}\right)^{-1.36}\right]e^{-1.14/\sigma^2} .
\end{equation}
To find the desired halo accretion rate, we then apply a method analogous to abundance matching \cite{Vale2004}, where we deduce the halo mass accretion rate by assuming that halos evolve at constant number density across redshifts. That is, we assume that the masses $(M_1, M_2)$ at $(z_1, z_2)$ are related via
\begin{equation}
    \int_{M_1}^\infty dM\, n_\mathrm{h}(M|z_1) = \int_{M_2}^\infty dM\, n_\mathrm{h}(M|z_2).
    \label{eq:ab-match}
\end{equation}
We obtain the accretion rate $\dot{M}_{\rm h}$ by numerically evaluating the time derivative of $M_{\rm h}$ in equation~\ref{eq:ab-match} under the limit $z_1 \rightarrow z_2$. We assume that the baryonic accretion rate, which is the source of star and dust formation, is simply proportional to the total accretion rate in most halos, except at high masses. In this picture, halos mostly grow through smooth accretion, as found by \cite{Goerdt2015} in numerical simulations, rather than mergers. The resulting growth histories are reasonably consistent with simulations \cite{Mirocha2021}, and the model is qualitatively similar to several others in the literature \cite{Trenti2010, Dayal2014, Mason2015}.

Both hydrodynamical simulation \cite{Keres2005, Birnboim2007, Faucher-Giguere2011} and empirically-motivated models \cite{Peng2010, Woo2013} have shown that baryonic accretion onto the central objects in high-mass halos is suppressed by their virial shocks, so we apply a shock suppression factor $f_\mathrm{sh}$ to the baryonic accretion rate given by the fit of \cite{Faucher-Giguere2011}. Thus the total accretion rate is $ f_{\rm sh} \dot{M}_b  \equiv f_{\rm sh} (\Omega_b/\Omega_m) \dot{M}_{\rm h}$.

In the minimalist model, we assume the baryonic accretion rate is in equilibrium with star formation and feedback, such that \cite{Dave2012}
\begin{equation}
    \dot{M}_b = \dot{M}_\star + \dot{M}_w,
\end{equation}
where $\dot{M}_\star$ is the total star formation rate, and $\dot{M}_w$ is the rate of baryonic expulsion due to feedback. As this feedback is generated by the stars, we assume that $\dot{M}_w = \eta \dot{M}_\star$, where the mass-loading parameter $\eta$ determines the strength of such feedback and is in general a function of halo mass and redshift. We also define the instantaneous star formation efficiency as the fraction of the accreted baryons that turn into stars, such that $f_\star \equiv \dot{M}_\star / \dot{M}_b$. Additionally, we impose an upper limit of star formation efficiency denoted $f_{\star,\mathrm{max}}$ to maintain the continuity of the resultant function. Our finalized star formation efficiency prescription therefore becomes
\begin{equation}
    f_\star = \frac{f_\mathrm{sh}}{f^{-1}_{\star,\mathrm{max}} + \eta(M_\mathrm{h}, z)}.
\end{equation}

We now require a prescription for the feedback parameter $\eta$. Our parameterization is based on two limiting scenarios. In the first, we assume that all of the kinetic energy from supernovae is available to drive the outflow. Therefore, we balance the supernova energy with the binding energy of the accreting gas \cite{Larson1974mar, Larson1974nov}, 
\begin{equation}
    \frac{1}{2}\dot{M}_w v^2_\mathrm{esc} = \dot{M}_\star \epsilon_k \omega_\mathrm{SN},
\end{equation}
where $v_\mathrm{esc}$ is the halo's escape velocity and $\omega_\mathrm{SN} \sim 10^{49} \mathrm{erg/M_\odot}$ is the fiducial supernova kinetic energy per unit mass of star formation, where the value here is typical of a Salpeter IMF. We leave the precise value free and fit it to galaxy measurements below as part of our overall normalization constant. This method assumes that the gas always receives a fixed portion of supernova blast energy. Alternatively, if most of the energy is lost to radiative cooling, we could obtain a similar relation by balancing the momentum of the supernova ejecta with the escape velocity \cite{Thompson2005, Krumholz2018}. In both cases, the feedback parameter can be fit by the general form
\begin{equation}
    \eta(M_\mathrm{h}, z) = A \left(\frac{10^{11.5}M_\odot}{M_\mathrm{h}}\right)^\xi \left(\frac{9}{1 + z}\right)^\sigma,
    \label{eq:eta}
\end{equation}
where $\xi$ and $\sigma$ are the power law indices and $A$ is the normalization factor. Here, we take $A=2.64$, $\xi=0.66$, and $\sigma=0.05$ from \cite{Hegde&Wyatt2024}'s MCMC fits. We remind the readers that these values are obtained by fitting to the pre-JWST $z=6-9$ UVLFs. 

To compare with existing observations, we convert the computed star formation rate of a galaxy to the corresponding UV luminosity it generates. Throughout this paper, we adopt the SFR-$L_\mathrm{UV}$ conversion from \cite{Madau2014}, which can be written
\begin{equation}
\dot{M}_\mathrm{\star} = \mathcal{K}_\mathrm{UV}\times L_\mathrm{UV},
\label{eq:SFRLUV}
\end{equation}
where the proportionality constant is $\mathcal{K}_\mathrm{UV} = 1.15\times10^{-28} \mathrm{M_\odot yr^{-1}/(erg\,s^{-1}Hz^{-1})}$ and $L_\mathrm{UV}$ is the continuum UV luminosity measured at 1500\, \AA. This conversion assumes a Salpeter IMF from 0.1 to 100~$M_\odot$ and an extended period of continuous star formation for $\gtrsim300 \ {\rm Myr}$.  

We have defined the star formation efficiency as an instantaneous value, but it will be useful to have the overall (time-averaged) value for a galaxy. We define $\Tilde{f}_\star\equiv M_\star/M_b$, where $M_b = (\Omega_b/\Omega_m) M_{\rm h}$. \cite{Furlanetto2017} showed that $\Tilde{f}_\star(M_\mathrm{halo}, z) \approx f_\star(M_\mathrm{halo}, z)/(1 + \xi)$, which we use as an estimate here; the slight error can be absorbed into our scaling constants. In the following subsections, we detail our dust model implementation based on \cite{Furlanetto2017}'s model. For clarity, we summarize our model's parameters in Table \ref{tab:f_tables} before introducing our implementation.
\begin{table}[tbp]
    \centering
    \begin{tabular}{|c|c|c|}
    \hline
    Symbol & Definition & Equation\\
    \hline
    $f_M$ & Mass scaling parameter & \ref{eq:fMdef}\\
    $f_{R}$ & Radius scaling parameter & \ref{eq:Rh}\\
    $f_d$ & Optical depth scaling parameter & \ref{eq:fd}\\
    \hline
    $P^{\rm esc}$& Galaxy-averaged escape probability for UV photons & \ref{eq:P_esc}\\
    $P^{\rm esc}_{\rm mix}$& Escape probability of the Homogenous mixture geometry & \ref{eq:P_mix}\\
    $P^{\rm esc}_{\rm shell}$& Escape probability of the shell geometry & \ref{eq:P_shell}\\
    $P^{\rm esc}_{\rm hol}$& Escape probability of the partially covered shell geometry& \ref{eq:pcov}\\
    \hline
    \end{tabular}
    \caption{Summary of scaling parameters used in this paper. The last column provides the equations in which these parameters are defined.}
    \label{tab:f_tables}
\end{table}

\subsection{Dust Production in High-Redshift Galaxies}

We next assume that the total dust mass is proportional to the total stellar mass with a yield $y_d$, such that the total dust mass is $y_d M_\star$. We will use a fiducial $y_d = 0.01$ value comparable to a typical disk galaxy from low redshifts. This choice of $y_d$ is in agreement with other modelling studies such as the semi-analytic model of \cite{Ferrara2022} (who found $\langle y_d\rangle\approx0.017$ using our notation), and the chemical evolution model of \cite{Calura2017, Millan-Irigoyen2020}. Similar values (0.007 and 0.003; calculated from product $y_d\nu$ from their paper) were also adopted by \cite{Dayal2022} in their dust model, ultimately motivated by the model of \cite{Todini2001, Bianchi2007} and also consistent with single SN observations in the local universe \cite{Temim2017, Rho2018}. Furthermore, the observational studies of \cite{Burgarella2020, Donevski2020, Witstok2023} found a similar dust yield range of $\sim0.001-0.01$. However, this $y_d$ parameter is uncertain at high redshifts, because the dust production mechanisms may differ, or a portion of the dust can be ejected by feedback \cite{Ziparo2023} or destroyed by supernova shock waves \cite{Cherchneff2010}. We therefore define a scaling parameter $f_M$ via
\begin{equation}
    M_\mathrm{dust} \equiv f_M \times (0.01 M_\star).
    \label{eq:fMdef}
\end{equation}
We naively expect $f_M \lesssim 1$, because dust formation may be less efficient at early times and because dust may be ejected or destroyed. For example, the disk models of \cite{Furlanetto2022} found that only  $\sim 0.1$ of the total baryonic mass associated with a halo was retained in the galaxy's interstellar medium, with most of the remainder ejected. We absorb those uncertainties (as well as any in the actual value of $y_d$ for these galaxies) into the mass scaling parameter $f_M$, which we will fix by comparison to observations in later sections. 

We note that we do not explicitly consider the effect of asymptotic giant branch (AGB) stars' dust production throughout this paper. Though AGB stars are the main sources of dust in local systems \cite{Matsuura2013} such as the Milky Way, \cite{Mancini2015} found that they are a less dominant factor compared to grain growth and SNe beyond $z\sim6$. Their cumulative effects could, however, also be absorbed into $f_M$.

\begin{figure}
    \centering
	\includegraphics[width=.5\textwidth]{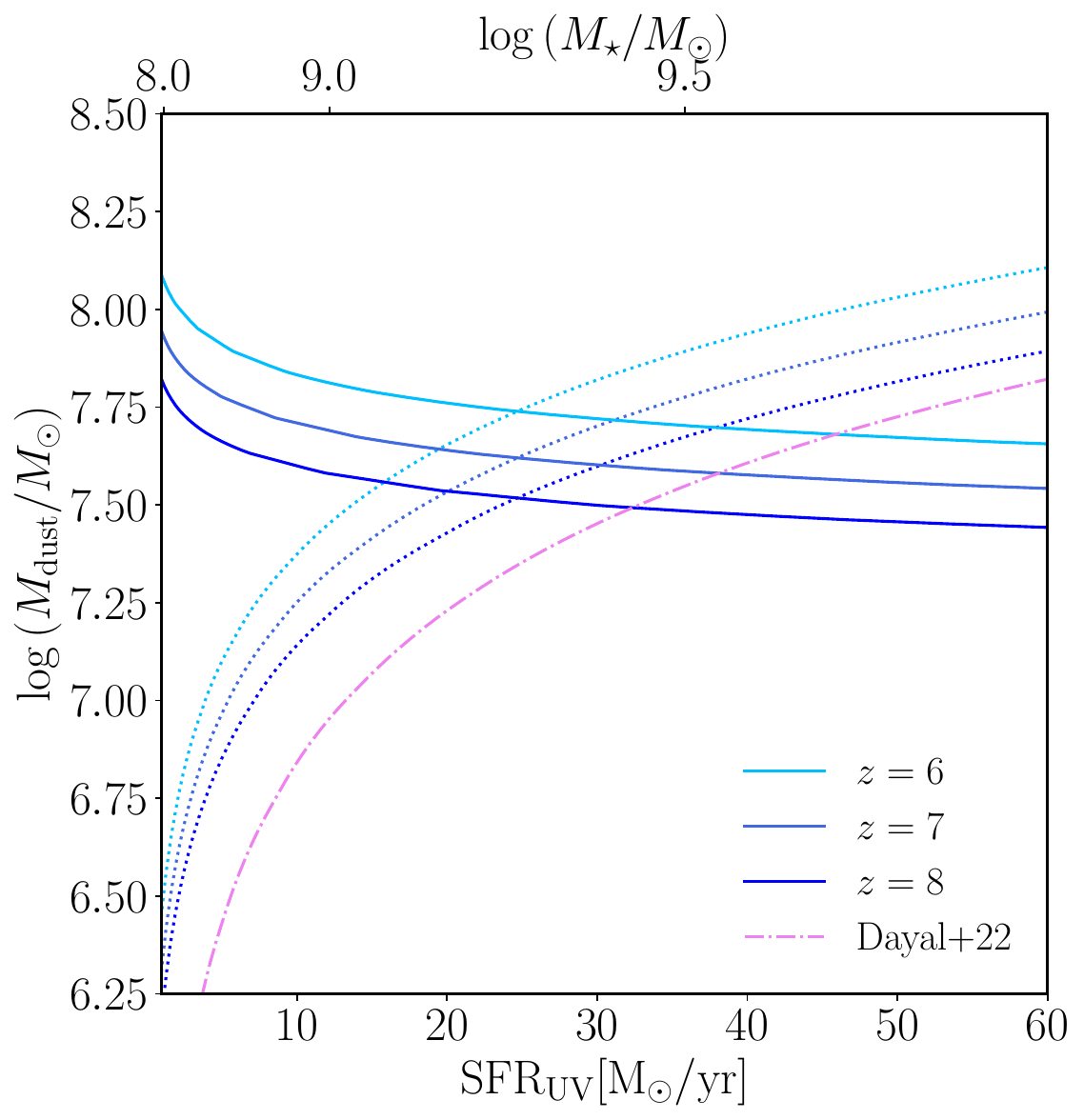}
    \caption{Expected dust masses as a function of SFR and $M_\star$ at $z=7$ (bottom and top axes; these are monotonically related in our model). The dotted curves show the dust mass assuming $f_M=1$, while the solid curves are the calibrated dust masses with $f_M$ formulated in section \ref{ALMAmethod}. The dash-dotted curve shows \cite{Dayal2022}'s dust masses at $z=7$.}
    \label{fig:DELPHI}
\end{figure}

The dotted curves in Figure~\ref{fig:DELPHI} show the resulting dust masses, assuming $f_M = 1$. We show them as a function of each galaxy's SFR, because observations are often presented that way. In our minimalist model, the SFR is a monotonic function of halo (and stellar) mass, so we provide the translation along the top axis. The solid curves preview the results of our model calibration to UV and IR observations -- we will find that $f_M$ may be a strong function of halo mass. The overall efficiency of dust formation is thus $\approx 0.01 f_M \tilde{f}_\star$, which for halos in the mass range $10^8$--$10^{12} \ M_\odot$ is typically $\sim 3 \times 10^{-4} f_M$. The expected dust mass is smaller in smaller halos, because their star formation efficiencies are small.

Before developing our model further, we briefly compare our predicted dust masses with \cite{Dayal2022}, who also use a semi-analytic dust treatment, albeit with more details in modelling specific dust evolution processes, such as dust grain growth, destruction, and astration. Their model uses a dark matter halo merger tree and traces the gas content associated with such halos to deduce the dust mass. In Figure \ref{fig:DELPHI}, we compare our work's fiducial dust mass (with $f_M=1$; dotted curves) to \cite{Dayal2022}'s results presented in their equation~(11) using our model derived $M_\star$ (shown in the top axis on the left panel). (In this comparison, we have converted the UV emissivity per unit star formation assumed by \cite{Dayal2022} to our value.) Our ``initial'' dust masses (with $f_M=1$) are typically a few times larger than their reported masses. However, the expected trends with halo mass are quite comparable. We do note that this discrepancy also increases with redshift: at $z \sim 10$, \cite{Dayal2022}'s model has dust masses more than an order of magnitude smaller than ours (if $f_M=1$). We return to this comparison in more detail in section~\ref{sec:comparison}.

\subsection{Dust Absorption}

We next use simple models (similar to \cite{Imara2018, Inoue2020}) to connect to observations through two observables: the absorption of UV photons and emission in the IR continuum. 

In the UV, the wavelength-dependent optical depth along a line of sight is
\begin{equation}
    \tau_\lambda = \int_0^{R_\mathrm{dust}}\rho_\mathrm{dust}\kappa_\mathrm{UV} dr,
    \label{eq:tau1}
\end{equation}
where $R_\mathrm{dust}$ is the dust radius, $\rho_\mathrm{dust}$ is the dust density, and $\kappa_\mathrm{UV}$ is the UV dust opacity.

The opacity depends on the choice of dust extinction law. As we focus on the rest-UV regime, we adopt a simplified power law approach from \cite{Mirocha2020}, with
\begin{equation}
    \kappa_{\rm UV} = \kappa_{1000}\left(\frac{\lambda}{10^3 \text{\AA}}\right)^{\gamma_k},
\end{equation}
where $\gamma_k = -1$ is the fiducial power law adopted in their work and $\kappa_{1000} = 10^5 \ \mathrm{cm^2 \ g^{-1}} \approx 20 \ \mathrm{pc^2 \ M_\odot^{-1}}$ is the dust opacity evaluated at 1000\,\AA. This prescription is consistent with the more sophisticated SMC dust extinction law presented by \cite{Weingartner2001}, which is a carbonaceous-silicate grain model (and which appears to match high-$z$ galaxies reasonably well given its steeper slope in the near-UV;  \cite{Bouwens2016}). We note that the simple power law estimate is reasonable for the UV continuum at 1500--1600\,\AA, but a more sophisticated dust extinction law may be warranted to consider the effects of specific atomic and molecular absorption lines (e.g. the 2175\,\AA\,``bump''), as described in detail by \cite{Draine2003}. To test the effect of particular extinction laws on our model, we have estimated the dust optical depth (equation~\ref{eq:tau1}) with both \cite{Calzetti2001} and \cite{Weingartner2001}'s results. We found that our model is not sensitive to the choice of extinction laws, thus motivating the choice of a simple power law. 

Because the integral in equation (\ref{eq:tau1}) extends over the entirety of the dust distribution, we need to model the dust's radial extent. We assume that the dust radius is proportional to the scale radius of a disk galaxy embedded in a dark matter halo. At high redshifts, the halo virial radius is by \cite{LoebFurlanetto2013} 
\begin{equation}
    R_{{\rm halo}} \approx 1.5\left(\frac{M_{\rm halo}}{10^8M_\odot
    }\right)^{1/3} \left(\frac{10}{1 + z}\right) \,  \mathrm{kpc}.
\end{equation}
Assuming the gas in a galaxy is rotationally supported, the galaxy scale radius is conventionally written as $R_{\rm gal} = (\lambda/\sqrt{2}) R_{\rm halo}$, where $\lambda$ is the spin parameter. We take $\lambda = 0.028$ as a fiducial value for a typical disk galaxy  \cite{Faucher-Giguere2018}. While this scale provides a reasonable estimate for the gas and stellar distributions, recent observational studies found that dust may be more extended than the UV-emitting star-forming regions \cite{Pozzi2024}. We therefore insert a radius scaling parameter $f_R$ 
\begin{equation}
    R_\mathrm{dust} = f_R  \frac{\lambda}{\sqrt{2}} R_\mathrm{halo},
    \label{eq:Rh}
\end{equation}
which we will later calibrate to observations. 

To evaluate equation~(\ref{eq:tau1}), we also need to determine the radial dust distribution, which is unconstrained. We will consider three simple morphologies below. But we note that, even after that distribution is specified, the optical depth for each star's light will vary depending on its location within the galaxy. We are ultimately interested in the dust's effects averaged over all lines of sight. Therefore, we describe this via the galaxy-averaged escape probability of UV photons $P^{\rm esc}$,
\begin{equation}
 L_{\rm obs} = P^{\rm esc} L_{\rm int},
 \label{eq:P_esc}
\end{equation}
where $L_{\rm int}$ is the intrinsic luminosity produced by the stars and $L_{\rm obs}$ is the apparent luminosity after dust processing for a distant observer. 

\subsubsection{Homogeneous Mixture}
We first consider a homogeneous, spherical mixture of stars and dust. This morphology corresponds to the scenario where dust is retained within the galaxy and uniformly mixed. In this case, equation~(\ref{eq:tau1}) becomes
\begin{equation}
    \tau_\mathrm{mix} = \frac{3M_\mathrm{dust}}{4\pi R^2_\mathrm{dust}}\kappa_\mathrm{UV}.
    \label{eq:tau_mix}
\end{equation}
Note that $\tau_{\rm mix} \propto f_M/f_R^2$
; if only the absorption is measured, the degeneracy between the dust mass and spatial distribution cannot be broken. We will fix this combination of parameters by comparison to UV observations in section~\ref{results}. We can then use additional observations -- dust emission in the IR -- to constrain the individual parameters in section~\ref{ALMAmethod}. 

For a homogenous mixture, \cite{Varosi1999} gives the escape probability of a UV photon as a function of optical depth as 
\begin{equation}
    P^{\rm esc}_\mathrm{mix}(\tau) = \frac{3}{4\tau}\left[1 - \frac{1}{2\tau^2} + \left(\frac{1}{\tau} + \frac{1}{2\tau^2}\right)e^{-2\tau}\right].
    \label{eq:P_mix}
\end{equation}

\subsubsection{Spherical Shell}
Next, we consider a spherical shell surrounding the galaxy. This may occur if feedback has ejected the dust to moderate distances from the galaxy (and the dust survives to these larger radii). In that case, equation~(\ref{eq:tau1}) becomes
\begin{equation}
    \tau_\mathrm{shell} = \frac{M_\mathrm{dust}}{4\pi R^2_\mathrm{dust}}\kappa_\mathrm{UV}.
    \label{eq:tau_shell}
\end{equation}
Note that this is three times less than $\tau_\mathrm{mix}$, as dust is farther from the stars (and hence more spread out) than in the mixture geometry. Note that we have the same $\tau_{\rm shell} \propto f_M/f_R^2$ relation. However, in this case, we have
\begin{equation}
    P^{\rm esc}_\mathrm{shell}(\tau) = e^{-\tau}.
    \label{eq:P_shell}
\end{equation}
Although the characteristic optical depth is smaller than the homogeneous case, the average escape probability of a shell decreases much more rapidly with optical depth, because the stars are all buried within the shell.

\subsubsection{Partially Covered Shell}
The shell model is motivated by galactic feedback ejecting dust from the central regions, but in reality, that process is highly anisotropic. We therefore next generalize the shell model to one with ``holes'' where UV photons can escape without any absorption. We define $f_\mathrm{cov}$ as the fraction of the galaxy's UV luminosity subject to dust absorption. (In a simple model, this would correspond to the fraction of lines of sight that intersect the dust, but because the stars may also be inhomogeneously distributed we leave the definition more open-ended.) This may happen if feedback is particularly strong in some directions and ejects the dust entirely out of the galaxy's host dark matter halo (or simply destroys it). The escape probability for UV photons in this geometry is then 
\begin{equation}
    P^{\rm esc}_\mathrm{hol} = (1 - f_\mathrm{cov}) + f_\mathrm{cov}[P^{\rm esc}_\mathrm{shell}(\tau_\mathrm{shell})],
    \label{eq:pcov}
\end{equation}
where $\tau_{\rm shell}$ is defined as in equation~(\ref{eq:tau_shell}).

Note that this geometry has an additional free parameter $f_{\rm cov}$ and will therefore require additional assumptions when we return to it in section \ref{ALMAfcov}.

\subsection{Dust Emission}
\label{Tdust}
Next, we estimate the IR emission from heated dust in order to compare it with ALMA observations. We will compare the observed continuum fluxes around a specific wavelength (near the [\cii] emission line at 158~$\mu$m, as that is a common window for observations). The total emission depends on the dust mass and temperature. We adopt an analytic temperature estimate from \cite{Inoue2020}, which assumes that the dust is in radiative equilibrium with the absorbed starlight and CMB heating. Then
\begin{equation}
    T_\mathrm{dust} = \left(\frac{L_\star^\mathrm{abs}}{C \kappa_\mathrm{IR,0} M_\mathrm{dust}} + T_\mathrm{CMB}^{\beta_\mathrm{IR} + 4}\right)^{1/(\beta_\mathrm{IR}+4)},
    \label{eq:T_dust}
\end{equation}
where the infrared absorption coefficient is given by a power law in frequency, $\kappa_{158}=\kappa_\mathrm{IR,0}(\nu/\nu_0)^{\beta_\mathrm{IR}}$ with $\kappa_\mathrm{IR,0} = 16 \ \mathrm{cm^2 \ g^{-1}}$, $\nu_0 = c/100 \ \mu {\rm m}$ is the normalization frequency, and $\beta_\mathrm{IR} = 1.6$ as the IR spectral slope. Both these fiducial values are taken from \cite{Bianchi2019}'s Table~1 and are appropriate for the Milky Way. Also, $T_\mathrm{CMB}$ is the CMB temperature at a given $z$. The constant $C$ is 
\begin{equation}
    C = \frac{8 \pi k_B^{\beta_\mathrm{IR}+4}}{c^2\nu_0h^{\beta_\mathrm{IR}+3}}\zeta(\beta_\mathrm{IR} + 4)\Gamma(\beta_\mathrm{IR} + 4), 
\end{equation}
where $k_B$ is the Boltzmann constant, $c$ is the speed of light, $h$ is Planck's constant, $\zeta$ is the Riemann zeta function and $\Gamma$ is the gamma function. The total absorbed UV stellar emission $L_\star^\mathrm{abs}$ is computed by integrating the absorbed UV luminosity across all frequencies, so
\begin{equation}
    L_\star^\mathrm{abs} = \int L_\mathrm{obs}\frac{1 - P^\mathrm{esc}(\tau_\lambda)}{P^\mathrm{esc}(\tau_\lambda)}d\lambda,
\end{equation}
where $\tau_\lambda$ is the wavelength-dependent optical depth. We normalize $\tau_\lambda$ with the effective optical depth at 1600\, \AA\, (which is uncertain until we fix the scaling parameters) and assume that $\tau_\lambda  \propto \lambda^{-1}$ at other UV wavelengths, as approximated by the \cite{Mirocha2020} extinction law. To compute this integral, we integrate from the Lyman break wavelength 912\,\AA\, to 4000\,\AA \ to cover the total energy released in the UV regime, which dominates energetically, though we find that the result is not sensitive to any choice for the upper limit beyond 2800\,\AA. We note that equation~(\ref{eq:T_dust}) includes heating from the CMB. Sometimes this effect is ignored in estimates taken from observations; in that case, one can take the limit $T_\mathrm{CMB} = 0 \  \mathrm{K}$ in equation~(\ref{eq:T_dust}).

This method assumes that the dust is at a single temperature, so that the absorbed starlight is distributed evenly across the dust grains. \cite{Ferrara2022} notes that when the dust is optically thick, grains will be exposed to a range of intensities so will reach a range of temperatures. They thus compute a luminosity-weighted temperature, such that the dust population closer to the source is hotter. We ignore this effect because our opacities (calculated from equations~\ref{eq:tau_mix}; \ref{eq:tau_shell}) are relatively small; we have found that it causes a $\lesssim 10\%$ temperature difference. However, such a correction may be more important in a different morphology or very optically thick medium. 

With the temperature in hand, we can next predict the far-IR continuum flux. To compare to the largest set of observations, we focus on the far-IR dust continuum flux around the [\cii] emission line -- it is likely the strongest line so is the target of many campaigns, and the dust continuum can be conveniently estimated elsewhere in the band. Hereafter, we use $F_{158}$ as a short-hand for the dust far-IR continuum flux in this window. To calculate $F_{158}$, we adopt the ``grey-body'' treatment of \cite{dacunha2013}, so
\begin{equation}
   F_{158} = \frac{1 + z}{d_L^2}\,M_\mathrm{dust}\kappa_{158} \left[B_{158}(T_\mathrm{dust}) - B_{158}(T_\mathrm{CMB}) \right].
   \label{eq:F_CII}
\end{equation}
where the $(1+z)$ factor results from the shift in the bandwidth, $d_L$ is the luminosity distance, and $B_{158}$ is the Planck function at the dust temperature at the emitted frequency $\nu$. We note again that some authors do not subtract the CMB contribution, so the last term may not be necessary when comparing to observations.

\section{Empirical Dust Attenuation Calibration}
\label{results}

Our simple dust model has two free parameters, characterizing the dust's spatial extent (through $f_R$) and its mass (parameterized by $f_M$). In this section, we will describe how we can use existing UV observations to place a joint constraint on these parameters. 

\subsection{UV Attenuation Calibration}

Ideally, one would estimate the UV dust attenuation via SED modelling, where the level of extinction is obtained in combination with the stellar parameters. However, such modelling is not yet available for large numbers of high-$z$ galaxies, so instead, we follow a common approximate method. This method assumes that the UV slopes $\beta$ of these galaxies depend on the dust properties (as dust reddens the spectrum). We use observed estimates of these slopes together with an empirical relation from \cite{Meurer1999} that relates the UV slope to the level of extinction. 

First, we require the UV slopes of ``typical'' high-$z$ galaxies as a function of magnitude and redshift, which we obtain from two observed samples detailed later. For simplicity, we assume that the average UV slope $\langle\beta\rangle$ is linear in the observed magnitude, with the slope $d \beta / d M_\mathrm{UV}$ and intercept $\beta_{M_0}$ varying linearly with redshift as well. Thus we have
 \begin{equation}
     \langle\beta\rangle \left(M_\mathrm{UV}^\mathrm{dust}, z\right) = \frac{d\beta}{d M^{\rm dust}_\mathrm{UV}}(z) \left[M_\mathrm{UV}^\mathrm{dust} - M_0\right] + \beta_{M_0}(z),
     \label{eq:beta}
 \end{equation}
 where $M_0$ is a reference magnitude, and we assume that
 \begin{align}
    \beta_\mathrm{M_0}(z) &= \frac{d\beta}{dz}z + \beta|_{z=0}, \\
    \frac{d\beta}{dM^\mathrm{dust}_\mathrm{UV}}(z) &= \frac{d}{dz}\left(\frac{d\beta}{dM^\mathrm{dust}_\mathrm{UV}}\right)z + \left.\frac{d\beta}{dM^\mathrm{dust}_\mathrm{UV}}\right|_{z=0}.
\end{align}

\begin{table}
	\centering
	\caption{The best-fit parameters of $\beta_\mathrm{M_0}$ and $ \frac{d\beta}{dM^\mathrm{dust}_\mathrm{UV}}$ to \cite{Bouwens2014} \& \cite{Topping2023a} as linear functions of redshift, displayed as purple curves in Figure \ref{fig:betafit}. The first column represents the slope of the linear function, and the second column represents the offset evaluated at $z=0$. }
	\label{tab:betafit}
	\begin{tabular}{lccr} % four columns, alignment for each
		\hline
		& slope & offset $(z=0)$ \\
		\hline
		$\beta_\mathrm{M_0}$ & $-0.081\pm0.016$ & $-1.58\pm0.115$\\
		$ d\beta/dM^\mathrm{dust}_\mathrm{UV}$ & $0.012\pm0.006$ & $-0.216\pm0.040$ \\
		\hline
	\end{tabular}
\end{table}

\begin{figure}
    \centering
	\includegraphics[width=.5\textwidth]{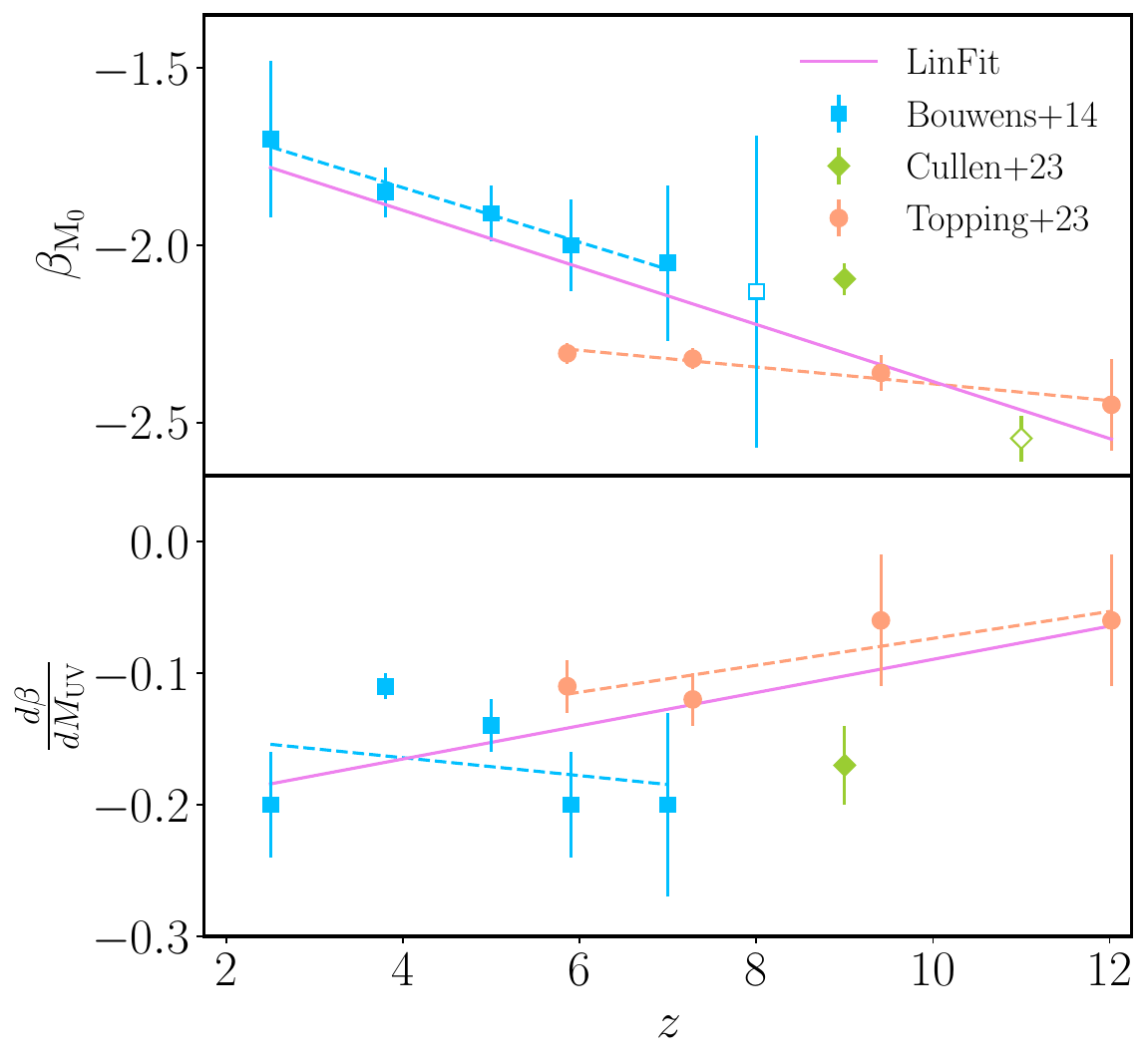}
    \caption{The redshift evolution of the UV slope fit parameters, $\beta_\mathrm{M_0}(z), \frac{d\beta}{dM^{\rm dust}_\mathrm{UV}}(z)$, measured from the \cite{Bouwens2014} (blue squares) and \cite{Topping2023a} (orange circles) data sets. We excluded \cite{Bouwens2014}'s measurement at $z=8$ (open square) from the fit since the corresponding slope was fixed. We also show measurements from \cite{Cullen2023a} at $\langle z\rangle\approx9$  and 11 (green diamonds; the latter was also estimated without varying the slope). These are not included in our fit. The purple curve exhibits the joint best fit to \cite{Bouwens2014} and \cite{Topping2023a}  with parameters presented in Table. \ref{tab:betafit}, whereas the dashed blue and orange curves are the best fit for the individual datasets. We align all the data to have $M_0=-19.5$.}
    \label{fig:betafit}
\end{figure}

To estimate these parameters, we use galaxies from \cite{Bouwens2014}, who measured slopes from over 4000 HST sources in deep surveys at $z=4$-$8$, and \cite{Topping2023a}, who measured slopes over $z=6$-$12$ with a large set of sources from the JWST Advanced Deep Extragalactic Survey (JADES). We summarize the fit parameters in Table. \ref{tab:betafit} and plot the resulting relation in Figure~\ref{fig:betafit}. We show the measurements at each redshift with the data points, and our best-fit linear relations for the parameters via the solid line. The dashed lines show the individual fits to the different datasets. Note that we have excluded the $z=8$ data point from \cite{Bouwens2014} for our fit, because its properties were estimated after holding the slope fixed. The qualitative evolution of the UV slopes is clear, although the precise form is only weakly constrained. 

Figure~\ref{fig:betafit} also includes measurements from \cite{Cullen2023a}, who find a steeper evolution in the zero-point than \cite{Topping2023a}. This is likely due to the sparseness of the data at $z\gtrsim10$. Given this, and the potential for contamination in photometric surveys \cite{Naidu2022, Furlanetto2023, Zavala2023} and cosmic variance \cite{Mason2023}, we do not include additional measurements. However, in Appendix~\ref{appendix}, we do compare our inferred relation to other recent JWST measurements, finding reasonable consistency (albeit with a large scatter amongst individual sources). 

We must now transform these UV slopes into estimates of the dust extinction. To do so, we use the IRX-$\beta$ relation from \cite{Meurer1999}, which is an empirical relation (calibrated to local starburst galaxies) that links the observed UV spectral slope $\beta$ to its corresponding infrared excess (IRX, defined as ${\rm IRX}=L_{\rm IR}/L_{\rm UV}$), which parameterizes the amount of dust emission (ultimately due to absorbed starlight). This can therefore be converted to the total UV emission in the absence of dust, $M_\mathrm{UV}^\mathrm{int}$ by \cite{Vogelsberger2020} 
\begin{equation}
M_\mathrm{UV}^\mathrm{int}= 
      \left(1-C_1\frac{d\beta}{d M_\mathrm{UV}^\mathrm{dust}}\right)M_\mathrm{UV}^\mathrm{dust} - C_0 - C_1\beta_{M_0} + C_1\frac{d\beta}{d M_\mathrm{UV}^\mathrm{dust}}M_0,
      \label{eq:vogelsberger}
\end{equation}
where we take constant values $C_0 = 4.43, C_1 = 1.99$ for absorption at 1600 \AA\, from \cite{Meurer1999}. We also require that $M_\mathrm{UV}^\mathrm{dust}\leq M_\mathrm{UV}^\mathrm{int}$ to prevent unphysical negative attenuation (which can happen as a result of our simple UV slope parameterization). This becomes important for faint galaxies with negligible dust production. We refer interested readers to the detailed derivation of equation~(\ref{eq:vogelsberger}) in \cite{Vogelsberger2020}'s section 3.2.1. Also, we note that we use luminosities evaluated at 1500 \AA, while the \cite{Meurer1999} relation uses the absorption at 1600 \AA. We convert between these with the power law $L_\lambda\propto (\lambda/\lambda_0)^\beta$. 

\begin{figure}
    \centering
	\includegraphics[width=.5\textwidth]{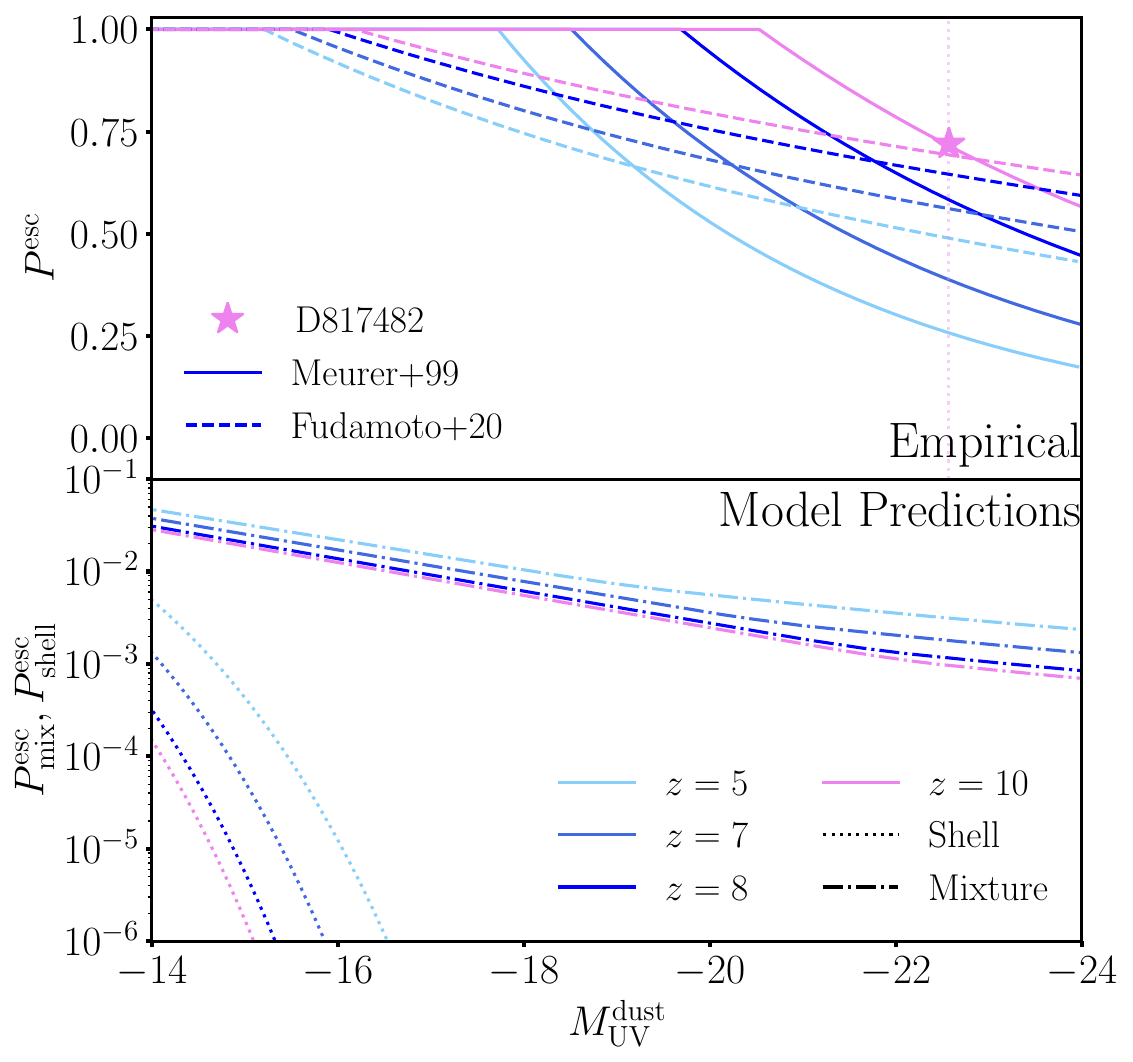}
    \caption{The escape probability of UV photons at 1600\,\AA\,in our model. The upper panel shows the \textit{empirical} probabilities calculated with the IRX-$\beta$ relations from \cite{Meurer1999} (solid curves) and \cite{Fudamoto2020} (dashed curves), respectively, whereas the lower panel shows the probabilities calculated \textit{from our models} assuming $f_d=1$ (equations~\ref{eq:P_mix}; \ref{eq:P_shell}). The violet star symbol represents the brightest galaxy in our JWST data compilation in Appendix \ref{appendix}, 817482 from the COSMOS field, from \cite{Cullen2023} at $z=9.89$. It has  $P_\mathrm{esc}\approx0.72$ according to the \cite{Meurer1999} relation at $z=10$.}
    \label{fig:Rabs}
\end{figure}

This procedure provides the intrinsic luminosity, and hence the probability that UV photons escape the system to the observer (via equation~\ref{eq:P_esc}). We show the results in Figure~\ref{fig:Rabs}. The attenuation increases towards lower redshifts and towards higher luminosities, as one would naturally expect. Interestingly, for much of the Cosmic Dawn, the attenuation is negligible except in very luminous galaxies: even at $z \sim 5$, $P_{\rm esc} \approx 1$ for $M_\mathrm{UV} \gtrsim -18$ \footnote{Hereafter, we use $M_{\rm UV}$ to denote the dust-processed or observed magnitude $M_{\rm UV}^{\rm dust}$ for conciseness unless otherwise noted.}, which corresponds to a halo mass $\sim 10^{11}M_\odot$. \footnote{This result motivates our temperature treatment in section \ref{Tdust}, where we assume a single mean dust temperature in the optically thin regime. }   As shown in appendix~\ref{appendix}, most observed systems at very high redshifts are near this limit, so the overall dust correction to the universal star formation rate density is expected to be modest during and before reionization (see section \ref{SFRD}). To provide some intuition, the star in Figure~\ref{fig:Rabs} corresponds to the brightest high-$z$ galaxy in the COSMOS field (see the Appendix), ID 817482, at $z\approx9.89$ \cite{Cullen2023}. In this case, the escape fraction of star formation is about $72\%$ according to our model. 

In the next subsection, we will compare these empirical results to our model. But before proceeding, we must note some caveats in adopting this empirical dust law. First, it is not clear that the IRX-$\beta$ relation of \cite{Meurer1999} can be applied to higher redshifts. Observations at $z\approx3$ have found consistency with \cite{Meurer1999} for faint galaxies, although more massive galaxies with more dust may present problems. At $z\approx5$, \cite{Fudamoto2020} finds the Infrared excess (IRX) in many galaxies to be smaller than that predicted by the \cite{Meurer1999} relation. One possible explanation is that the dust and star-forming regions may be spatially segregated in the high-$z$ regime so that the dust is illuminated by only a fraction of the UV light \cite{Ferrara2022, Inami2022, Algera2023}. As the IRX-$\beta$ relation assumes the UV and IR emitting regions to be co-spatial, extrapolation from low-$z$ may result in an overestimation of dust reddening. We return to a detailed discussion about the current IRX-$\beta$ relations in section \ref{IRX}.

To test the importance of the particular IRX-$\beta$ relation, we have also estimated the escape probability using the empirical relations from \cite{Casey2014, McLure2018, Reddy2018, Fudamoto2020}. In Figure~\ref{fig:Rabs} we also show the photon escape fractions inferred from the IRX-$\beta$ relation measured by \cite{Fudamoto2020} from a (small) sample of higher redshift galaxies ($z\sim5$). The results are qualitatively similar to our fiducial model, but they show a gentler dependence on halo mass (or luminosity). Given that these faint galaxies are only now beginning to be observed by JWST, we stick with the widely-used \cite{Meurer1999} relation for consistency with past work. However, for global properties where the faint galaxies play a significant role (e.g. SFRD; section. \ref{SFRD}), we will also show an estimate from \cite{Fudamoto2020}'s IRX-$\beta$ relation.

\subsection{Inferred Dust Properties}

With the empirically-derived dust attenuation in hand, we can now begin to calibrate the free parameters in our model. We will focus on the shell and mixture morphologies for now, which have parameters for the dust spatial extent ($f_R$, defined relative to the disk scale length) and dust mass ($f_M$, defined relative to the total amount of dust expected to be created). However, because the escape fraction only depends on the dust optical depth in these simple models, we can only constrain their combination 
\begin{equation}
    f_d \equiv \frac{f_M}{f_R^2},
    \label{eq:fd}
\end{equation} 
with the UV properties. 

The lower panel of Figure~\ref{fig:Rabs} shows the predicted UV escape fractions in our shell and mixture models assuming $f_d=1$. It is immediately obvious that the models predict far more absorption than is observed, with the model optical depths one or two orders of magnitude larger than the empirical results. This implies that either the dust is more spread out (so that each line of sight has a smaller optical depth) or that the galaxies have less dust than naively expected (or both). This appears to be a generic conclusion of galaxy formation models at these redshifts (e.g., \cite{Mirocha2020, Dayal2022, Palla2023, Ziparo2023}). Note that our model also predicts an opposite redshift-evolution trend compared to that of the observations, with the optical depth increasing with redshift in the models. This is because star formation (at a fixed halo mass) is slightly more efficient at high redshifts and (more importantly) because galaxies are more compact. 

\begin{figure}
    \centering
	\includegraphics[width=.5\textwidth]{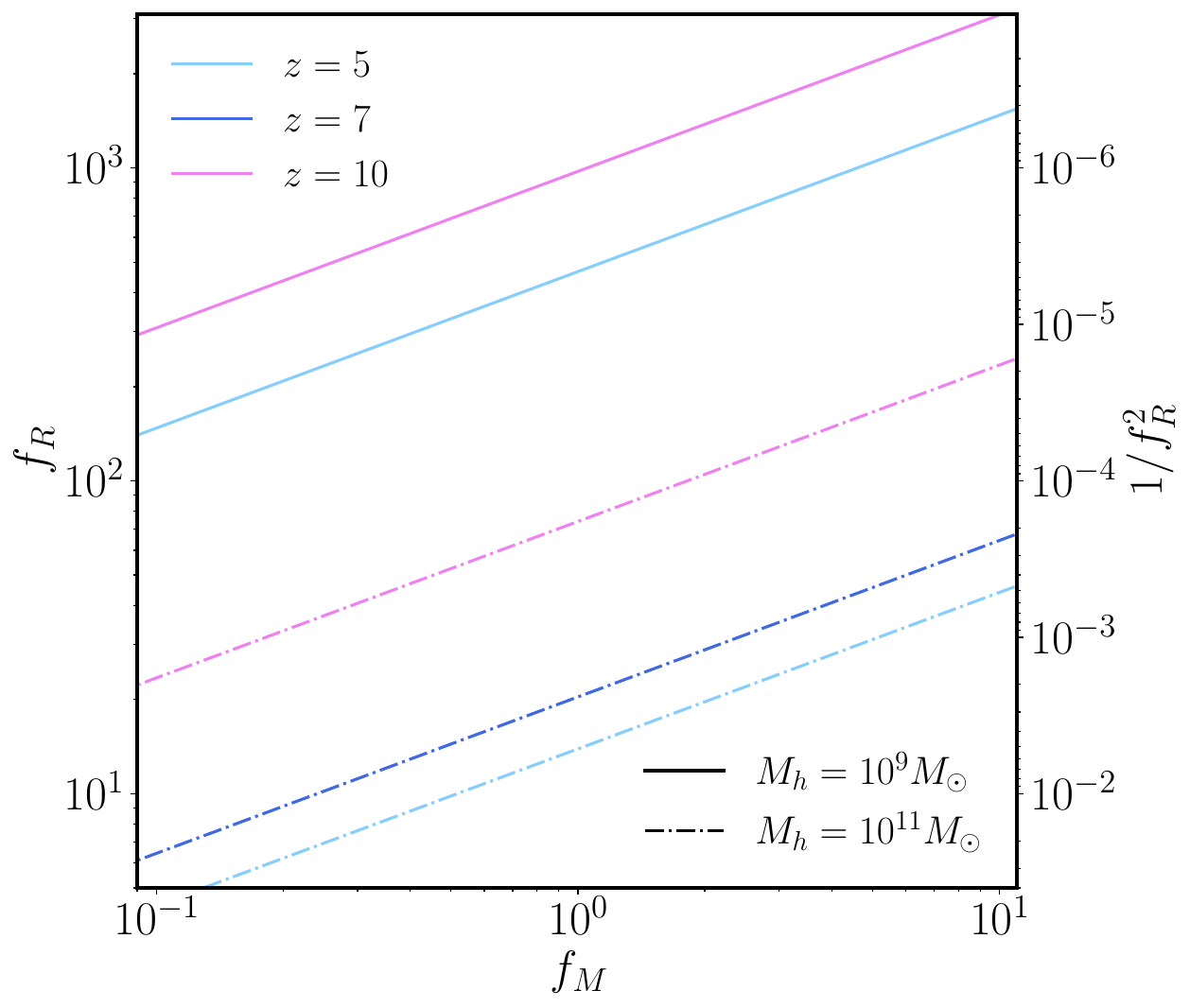} 
    \caption{Contours showing the loci of possible solutions $(f_M, \, f_R)$ that gives the empirically-derived $f_d$ across the modelled galaxy populations by our UV calibration at various redshifts. The solid curves are computed for a halo mass $10^9 M_\odot$ and the dash-dotted curves take $10^{11} M_\odot$. All spaces above each contour are also possible solutions for galaxies that require a smaller $f_d$.}
    \label{fig:fdust}
\end{figure}

Figure~\ref{fig:fdust} shows the loci of $f_d$ values that bring our model in agreement with the UV observations at a variety of redshifts (assuming that the IRX-$\beta$ relation from \cite{Meurer1999} is accurate). As shown, the required $f_d$ value decreases towards higher redshifts and lower halo masses as dust extinction decreases. Because the mass and radius scaling parameters are degenerate, a range of solutions is possible at any given redshift. To isolate their values, we require additional observations of the dust emission, which we will explore in the next section.

\section{Constraining the Model with Dust Emission}
\label{ALMA}
With its ability to detect the rest-frame far-IR dust continuum of high-$z$ galaxies, the Atacama Large Millimetre Array (ALMA) has become our heaviest artillery for understanding dust in early cosmic times. Specifically, ALMA surveys targeting the [\cii] line can measure (or constrain) the nearby dust continuum. In this section, we will use the IR emission to break the degeneracy between the dust parameters.

\subsection{The ALMA Comparison Sample}
We focus on two ALMA surveys: the ALMA Large Program to INvestigate [\cii] at Early times \cite[ALPINE,][]{Bethermin2020} carried out in Cycle-5 and Cycle-6, and the Reionization-Era Bright Emission Line Survey \cite[REBELS,][]{Bouwens2022} carried out in Cycle-7. With the shared objective of providing large statistical samples of far-IR continuum observations of UV-selected bright galaxies, ALPINE observed the [\cii] continuum emission of 23 galaxies in the redshift range $4.4\lesssim z\lesssim5.9$, while REBELS observed 14 galaxies with a mean redshift around $\langle z \rangle\approx7$. Complementary studies from both surveys supplement UV and optical properties of the observed galaxy populations, presented by \cite{Faisst2020} for ALPINE and  \cite{Bouwens2022} and \cite{Ferrara2022} for REBELS, respectively. 

Although the analyses of these surveys present many properties of the observed galaxies, we must be careful comparing them to our model. For example, the star formation rate of these galaxies inferred from their UV luminosity depends on assumptions about the IMF, metallicity, etc., while the reported stellar masses depend on the details of the SED modelling. We will generally report our results as a function of the SFR, because it is more easily related to observables. In order to ensure a meaningful comparison, we adopt the conversion in equation~(\ref{eq:SFRLUV}), applied to the reported luminosities. For ALPINE, we convert the UV magnitudes at 1500\, \AA\, from \cite{Faisst2020} to the SFR. For REBELS, we convert \cite{Ferrara2022}'s total SFR to the observed-UV SFR using their derived transmissivity factor $T_\mathrm{1500}$, while also replacing the SFR-$L_\mathrm{UV}$ conversion factor in their equation~(1) with our choice for $\mathcal{K}_{\rm UV}$.

One additional complexity is that stellar mass inferences depend on assumptions about the stellar populations and on the photometric library used for the measurement. The assumptions of these codes typically do not match the results of high-$z$ galaxy models (e.g., \cite{Mirocha2023}). The SED-fitting process is complex enough that we cannot easily translate between the inferences and models, so we will not attempt to do so here. Instead, we will mostly focus on comparisons as a function of SFR, which are at least transparent to convert. When we report stellar masses, we will simply use the results of the ALPINE and REBELS groups without modification.

However, we note that our minimalist model assumes a one-to-one correspondence between halo mass, stellar mass, and SFR. Real galaxies are of course variable, both because of the differing halo environments and because of temporal variations in individual systems. Because the stellar mass is a cumulative measure, it is relatively stable when averaging over this variability  -- while the SFR against which we will compare is much more sensitive to it. There are indications that galaxies at $z \gtrsim 8$ are extremely bursty \cite{Mason2023, Mirocha2023, Munoz2023}, so we will return to this issue in section~\ref{bursts}. For now, we note that we should not take the comparison too seriously for any individual system. 

\begin{figure}
    \centering
	\includegraphics[width=.5\textwidth]{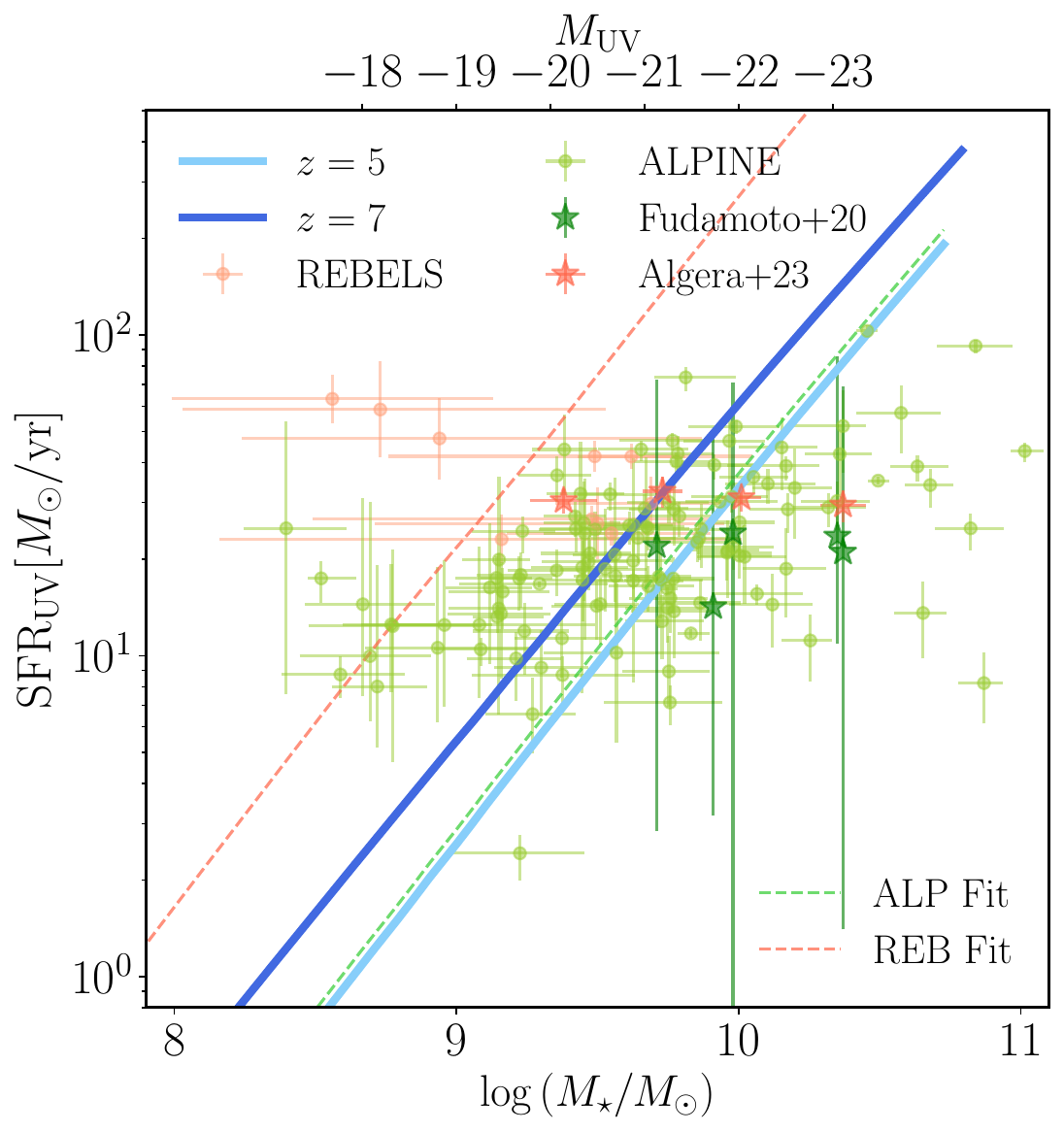}
    \caption{The star-forming main sequence at $z=5$ and 7 as seen in our models (solid curves) and the ALMA surveys. The orange points are from the REBELS survey \cite{Ferrara2022} at $z \approx 7$, corrected to our assumptions about the stellar populations. The green points are from the ALPINE survey \cite{Faisst2020} at $z\approx4.5$--5.5. The dashed curves show fits to these data sets if we constrain the slope to follow that expected in the model. The darker star symbols show results from stacking analyses that include sources undetected by these surveys \cite{Fudamoto2020, Algera2023}. We also show the conversion between stellar mass and UV magnitude in our model along the top axis. }
    \label{fig:Mstar_SFR_reb}
\end{figure}

Figure~\ref{fig:Mstar_SFR_reb} provides a rough comparison between the model and observed samples in terms of their ``star-forming main sequence'' relating SFR and stellar mass. The model results are shown by the solid curves, while the individual galaxies in the surveys are shown by the circles with error bars. Visually, the ALPINE data are more or less consistent with the model, but the REBELS points lay significantly above the models. This is confirmed by the dashed curves: these are fit to the data points, assuming that they follow the same power-law trend as the model but varying the normalization. The REBELS fit is driven higher by the three leftmost points, which have large UV luminosities (and hence SFRs) but small inferred stellar masses. These could be vigorous starbursts in small galaxies, or they could have ``hidden'' mass components from old stellar populations. (We note that the SED-based stellar masses did not have JWST data available, so are mostly insensitive to such old stars.)

%%%%%%%%% FIGURE: ALMA SF main sequence

However, the individual survey detections are biased, as they only include galaxies with bright enough dust emission to be detected. Both surveys also have significant numbers of undetected sources. The population properties must, therefore, be constrained through stacking analyses that include these fainter sources. The dark stars in Figure~\ref{fig:Mstar_SFR_reb} show the results of 
such analyses for both surveys \cite{Fudamoto2020, Algera2023}.
Again, ALPINE is reasonably consistent with our model, albeit over a limited range in stellar masses, but we see that now REBELS is also in reasonable agreement. We also note that both surveys find a much flatter relationship between $M_\star$ and SFR than our model provides; whether this is a selection effect is unclear, given the limited dynamic range in stellar mass.

\subsection{Implications for the Dust Properties}
\label{ALMAmethod}
We are now able to leverage the ALMA dust emission measurements to further constrain our models. In section \ref{results}, we used the UV attenuation inferred from the IRX-$\beta$ relation to estimate the overall optical depth, and from that, we placed a joint constraint on the mass and radius scaling factors. The dust emission also depends on the dust mass, so adding it can break the degeneracy between mass and radius. 

%%%%%%%%% FIGURE: dust measurements
\begin{figure*}
    \centering
	\includegraphics[width=.95\textwidth]{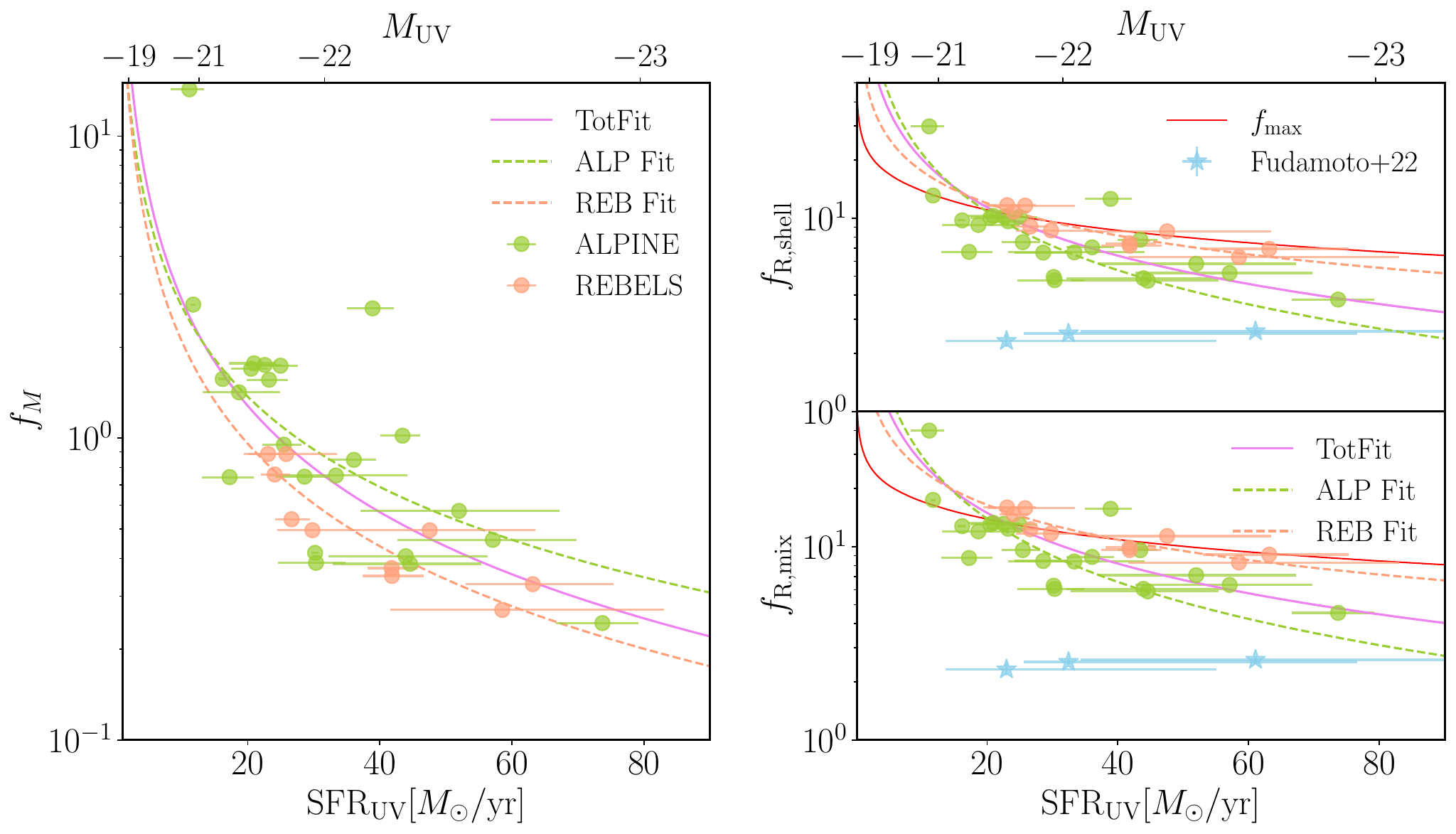}
    \caption{Dust scaling factors in our model, inferred through the combination of UV attenuation and dust emission. The green points correspond to ALPINE galaxies (using the dust continua reported by \cite{Bethermin2020}, while the orange points are from the REBELS survey \cite{Ferrara2022}. The dashed curves show power-law fits to each survey's results, while the solid curve shows a power-law fit to their combination. \emph{Left:} The mass scaling parameter $f_M$. \emph{Right:} The radius scaling parameters in the shell (top) and mixture (bottom) morphologies. In each of these panels, we also show the resolution of the individual REBELS detections as the solid red curve ($\approx 6.3 \mathrm{kpc}$) and the estimated sizes from stacking sources from both surveys with stars \cite{Fudamoto2022}. }
    \label{fig:fmfr}
\end{figure*}

Figure~\ref{fig:fmfr} shows our main results. The left panel shows the inferred value of $f_M$ for each galaxy (which is independent of the assumed dust morphology in the context of our simple model), while the right panels show the inferred values of $f_R$ for the shell and mixed dust morphologies. Interestingly, we find that relatively high dust masses -- with $f_M \sim 1$ -- are required to produce the overall dust luminosity. This implies that the systems are able to retain most of the dust they produce. However, we have already seen that the optical depth is much lower than expected in this case, so we would also need $f_R \sim 10$ for each model. The physical dust radius would then be $\sim f_R \lambda R_{\rm vir} \sim 0.3 R_{\rm vir}$. Thus, the dust would mostly be mixed through the circumgalactic media of these halos, assuming the spherically symmetrical geometries hold. This requires the dust to survive after being ejected from its formation regions, which may be challenging (though dusty clouds in outflows have been observed; \cite{Bland-Hawthorn2003, Rupke2013, Triani2021, Katsioli2023}).

We also see some potential trends in the data. The REBELS points tend to have slightly smaller $f_M$ and larger $f_R$ than the ALPINE points, but the differences are small and within the scatter of the points at each redshift. For a simple estimate, we therefore take a redshift-independent average fit. 

The trends with SFR are more obvious. In contrast to intuition from feedback-based models, $f_M$ appears to \emph{decrease} as the star formation rate increases (as previously found by \cite{Dayal2022, Palla2023}). If we follow the minimalist galaxy formation model and assume that the SFR increases monotonically with halo mass, this means that small halos would be more likely to retain their metals than massive halos -- contrary to the expectations of galaxy evolution models based on feedback. One possibility is enhanced dust destruction in massive systems, possibly due to the increasing amount of SNe shocks \cite{Cherchneff2010}. On the other hand, $f_R$ also decreases with SFR -- so that the dust that remains is closer to the galaxy's central regions in more massive systems. 

We must be cautious in interpreting these trends however, as we would expect similar qualitative results from burstiness: galaxies that have undergone recent starburst episodes that have now faded (so lay at small SFRs) may have residual dust from their previous event (so appear to have a large $f_M$). Going forward, we therefore consider two cases: one in which we take a single average value for each parameter (mass- and redshift-independent) and one in which we assume a power-law dependence on SFR. Fits for the latter are shown in Figure~\ref{fig:fmfr}. The redshift-independent fit (where we vary both the normalization and power-law slope) is given by 
\begin{equation}
\label{eq:fM}
f_M(\dot{m}_\star)=42.4\,\left(\frac{\dot{m}_\star}{M_\odot/{\rm yr}}\right)^{-1.17},
\end{equation} 
If we instead assume a constant value, we find the median $f_M$ value across the galaxy sample to be $\approx0.75$. 

\begin{figure}
    \centering
	\includegraphics[width=.5\textwidth]{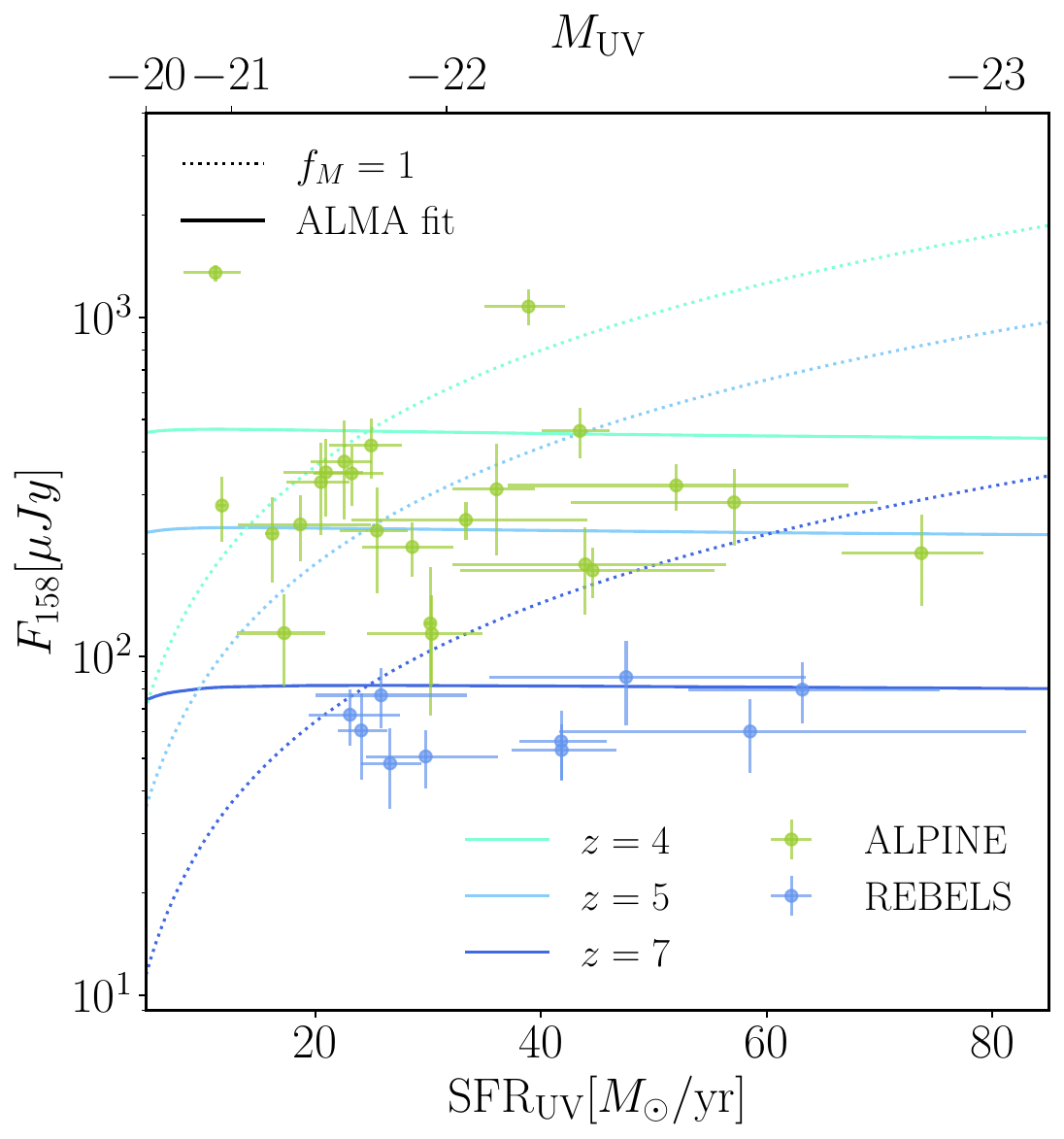}
    \caption{Comparison of model and observed dust continuum fluxes for the ALPINE and REBELS galaxies. The solid curves show the model fluxes using the power law fit of equation~\ref{eq:fM}, while the dashed curves take $f_M=1$. The green points are observed by ALPINE with $4.4\lesssim z\lesssim5.9$, while the blue points are observed by REBELS with $\langle z \rangle \approx 7$.}
    \label{fig:FCII}
\end{figure}

These simple dust geometries evidently require that the dust be an order of magnitude more spread out than one might naively expect. However, further studies from ALPINE and REBELS collaborations indicate that the dust radii are only $2-3$ times larger than that of the UV radii \cite{Fudamoto2022, Pozzi2024}. \cite{Inami2022} found that the REBELS galaxies typically have dust radii below the ALMA resolution, which was approximately $R_\mathrm{max}\approx 6.3 \, \mathrm{kpc}$ for their survey  (shown by the red curve in the right panels of Figure~\ref{fig:FCII}). The radii required by our models are dangerously close to this limit, with some even above it. Moreover, \cite{Fudamoto2022} have measured the extent of the dust emission in these sources with a stacking analysis, shown by the stars in Figure~\ref{fig:fmfr}. Stacking UV and dust emission is quite challenging, as it requires the sources to be spatially aligned, whereas the dust is often offset (by an unpredictable amount) relative to the UV emission in any given source \cite{Ferrara2022, Inami2022}. Nevertheless, the observed dust radii are several times smaller than expected in our model. This motivates a more complicated geometry that we consider in the next section. 

%%%%%%%%%%%FIGURE: Dust Flux

Before turning to that, however, we first examine a couple of other aspects of the model framework. Figure~\ref{fig:FCII} compares the resulting model fluxes (solid lines) to the observations (points with error bars). Because we have taken a redshift-independent fit, the model underestimates the ALPINE sources and overestimates those of REBELS sources by a modest amount. This also illustrates another potential limitation of the surveys: because they are limited by the survey depths, they cannot determine how wide the distribution of dust luminosities is for a given SFR. The dotted lines show our model with $f_M=1$, which decreases at small SFR -- unfortunately, the surveys would not be sensitive to such a decline. 

%%%%%%%%%%%FIGURE: Dust Temperature
\begin{figure}
    \centering
	\includegraphics[width=.5\textwidth]{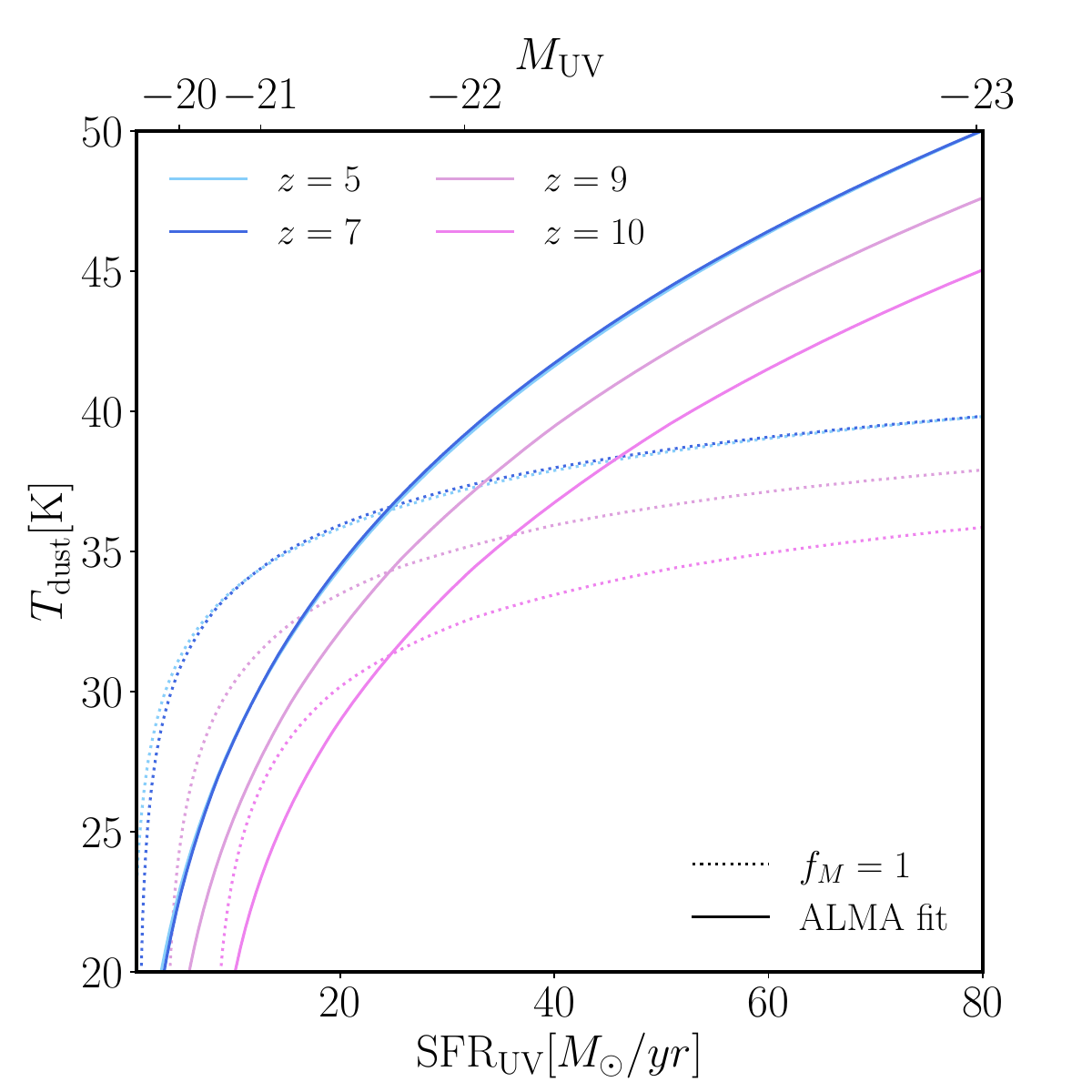}
    \caption{Dust temperatures (neglecting the CMB) in our model. The solid curves show the ALMA-calibrated temperature with the $f_M$ fit provided by equation~\ref{eq:fM}, while the dotted curves show the temperatures assuming $f_M=1$. }
    \label{fig:Tdust}
\end{figure} 

As described above, our model also estimates the dust temperature. This has not been directly measured for these sources, because they have only been observed over a very narrow range of wavelengths, but we show the model results in Figure~\ref{fig:Tdust}; here we take $T_{\rm CMB}=0 \ {\rm K}$ in equation~(\ref{eq:T_dust}), as often done when interpreting observations. The true dust temperatures are somewhat smaller, because the CMB contributes to the emission as well. The solid curves show our model with SFR-dependent fits, while the dotted curves are with constant parameters. Since $f_M$ decreases towards large ${\rm SFR}$, the calibrated dust temperature is higher than the original on the luminous end and lower on the faint end. However, for the dominant luminosity range in the ALMA surveys ($-21\lesssim M_{\rm UV}\lesssim-19$, or equivalently ${\rm SFR_{UV}}\lesssim40 \ {\rm M_\odot/yr}$), the calibration does not cause a significant difference in dust temperature.

\subsection{Gaps in the Dust Distribution?}
\label{ALMAfcov}
In the previous subsection, we found that our model required quite extended dust distributions in order to account simultaneously for the large dust luminosity and modest extinction. This occurs in both the limiting cases of dust mixed uniformly with the stars and dust expelled to a shell outside of the central galaxy, but it is in tension with observations \cite{Fudamoto2022, Inami2022} of the radius of the dust emitting region. One potential resolution is to drop the assumption of spherical symmetry in both our shell and mixture morphologies: we now turn to the partially covered shell model, in which dust appears only along a fraction $f_{\rm cov}$ of the lines of sight. 

Such a model does seem reasonable in light of other observations. \cite{Inami2022} also found a spatial offset between the UV and IR emission peaks in the REBELS survey, suggesting along with \cite{Algera2023}'s stacking analysis that the UV and IR emitting regions may not be co-spatial. \cite{Topping2022} concludes the same as a part of their specific SFR analysis joining UV and dust emission. In our ``partial shell'' model, this could be interpreted as a shell with a small covering factor and large radius, so that the dust region is confined to a small area displaced from the stars. More generally, this could occur if dust coincides with only a fraction of the star-forming regions. 

%%%%%%%%%%%%% FIGURE: covering fraction
\begin{figure}
    \centering
	\includegraphics[width=.5\textwidth]{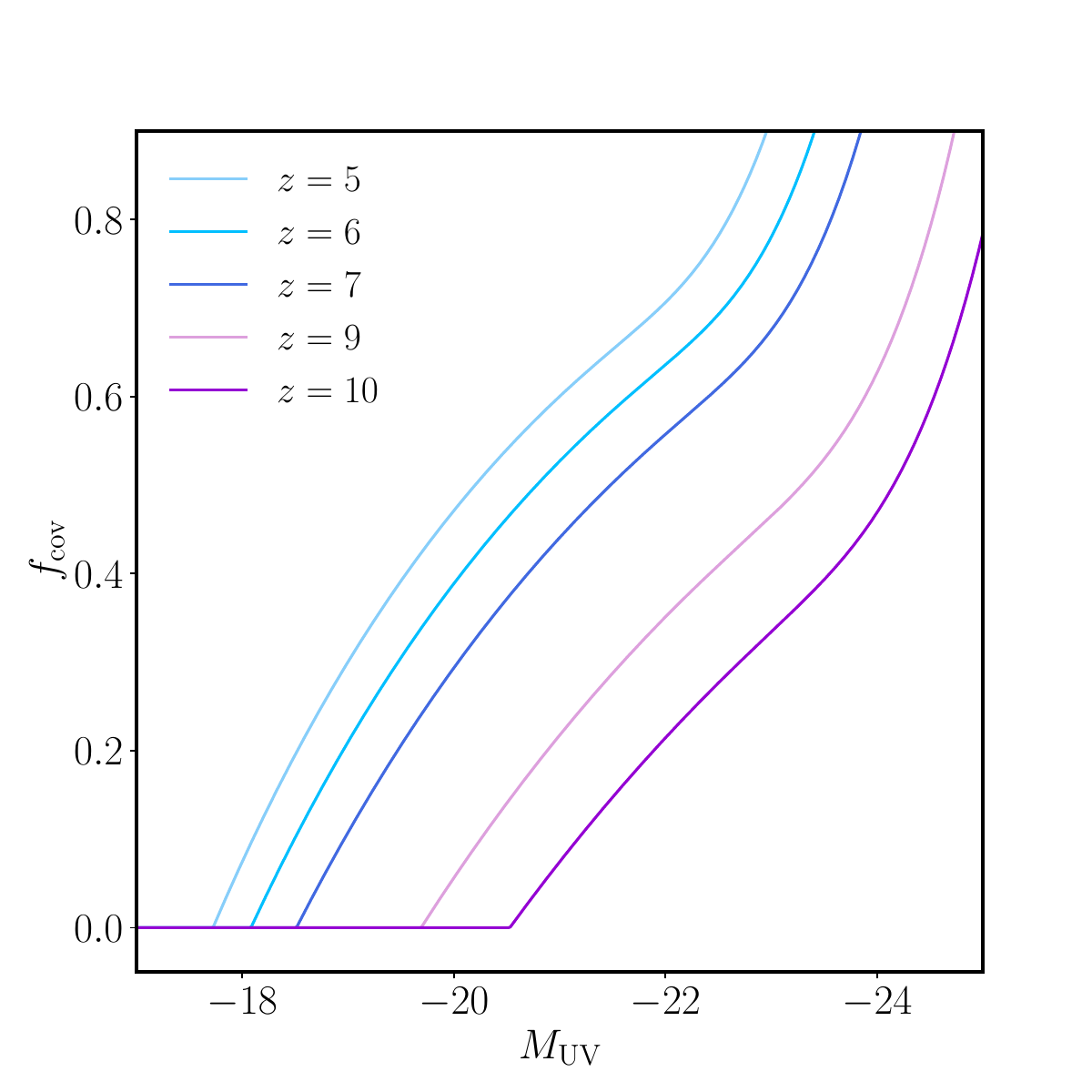}
    \caption{The covering fraction $f_\mathrm{cov}$ as a function of UV magnitude at various redshifts in our partially covered shell model. The curves are calculated from a scenario with the median $f_M$ value of 0.75 from the previous fit and a fiducial scaling $f_R=3$, motivated by \cite{Fudamoto2022}'s stacking analysis.}
    \label{fig:fcov}
\end{figure}

However, a morphology with ``holes'' introduces a new free parameter into the system, which is hard to constrain given the existing observations. We therefore take a simplified model in which: \emph{(i)} we take the median $f_M$ value of $0.75$ from all available $f_M$ values from both surveys (all data points in the left panel Figure~\ref{fig:fmfr}) and \emph{(ii)} we set $f_R=3$ to match the stacked measurements of  \cite{Fudamoto2022}. We then fix the covering fraction by equating the left side of equation~(\ref{eq:pcov}) to the empirically-determined UV escape probability. We assume that all these parameters are mass-independent. Note that the covering fraction does not affect the dust emission, because we hold its mass fixed (see equation~\ref{eq:tau_shell}), implicitly assuming that the dust remains in the system but is confined to a smaller region. Our results are not terribly sensitive to these choices, so long as we use values close to those suggested by the previous section. 

We show the resulting covering fractions in Figure~\ref{fig:fcov}. The small dust attenuation in faint galaxies implies that they are nearly clear of dust, but the covering fraction becomes large in luminous systems. More luminous galaxies have $0.2\lesssim f_{\rm cov}\lesssim0.7$, implying that the dust is more uniformly distributed in such systems. It seems reasonable that their longer star formation histories and larger SFRs mean less stochastic feedback and more time for metal mixing. Conversely, dust in smaller galaxies could be heavily impacted by bursty feedback, resulting in ``holes'' in the feedback-affected region and dust clumps in the unaffected region.

\section{Discussion}
\label{further}

Here we examine some other aspects of the dust model and its consequences.

\subsection{Comparison to Past Work}
\label{sec:comparison}

Other groups have also considered the dust content of high-$z$ galaxies. Here we compare to a few particularly relevant studies. \cite{Dayal2022} used their DELPHI semi-analytic model to estimate dust properties. They calibrated their star formation parameters to the observed UVLF and stellar mass function (using HST surveys). They then have physically motivated prescriptions for dust production, destruction, and escape via feedback, following its abundance in each system over time; while their dust production parameters are comparable to ours, we do not explicitly model destruction or escape. Figure~\ref{fig:DELPHI} compares our dust mass estimates to theirs (dash-dotted line). By comparing to ours, we find that their model corresponds to an effective $f_M \sim 0.3$. They compare to REBELS sources and find that their model underpredicts the dust mass inferred from infrared emission measurements, which we also find for the dominant low-mass galaxy population. Similarly, \cite{Palla2023}  find that their own semi-analytic dust model underpredicts the emission, arguing that the tension could be alleviated by a top-heavy IMF (amongst other possibilities).

\cite{Ferrara2022} and \cite{Sommovigo2022} have also estimated dust masses and temperatures from the REBELS sample. \cite{Ferrara2022} uses more sophisticated dust distributions and radiative transfer models, including a disk geometry and allowing for the dust to be at a range of temperatures. They use the $\beta$ slope and stellar population synthesis to break the degeneracy between dust mass and radius. Interestingly, their dust radii derived from $F_{158}$ are consistently lower than ours but their dust masses are consistent with ours. Thus, like our model and those of  \cite{Dayal2022, Palla2023}, they also find a high IR-to-UV ratio, concluding that this could possibly suggest a non-uniform dust morphology. To compare their results to our temperatures, we convert their luminosity-weighted temperature ($T_d'$) in Table 2 back to the mean temperature without the CMB effect ($\bar{T}_d$) described by their equation (15). (We remind the reader that our model finds that most high-$z$ galaxies are in the optically thin regime, so this has only a modest effect.) After this conversion, \cite{Ferrara2022} finds temperatures of $30$--$60 \ \mathrm{K}$ in the observed sources. Meanwhile, \cite{Sommovigo2022} find $40$--$58\ \mathrm{K}$. Even with rather different models, the predicted temperatures all roughly agree; this is not too surprising given that the dust luminosity is so sensitive to temperature.  

\cite{deRossi2023} used an even simpler approach than ours to estimating dust continuum fluxes, using a fixed star formation efficiency, assuming that all halo gas is retained, and imposing a fixed metallicity and dust-to-metal ratio. They take a much more sophisticated model for the dust properties, however, and then predict the far IR luminosities of very high-$z$ sources. Their fiducial model is comparable to ours, though they show that the dust emission can vary significantly depending on their assumptions about its microscopic properties. 

\subsection{The IRX-$\beta$ Relation}
\label{IRX}

%%%%%%%%% FIGURE: Variations in IRX-beta
\begin{figure}
    \centering
	\includegraphics[width=0.5\textwidth]{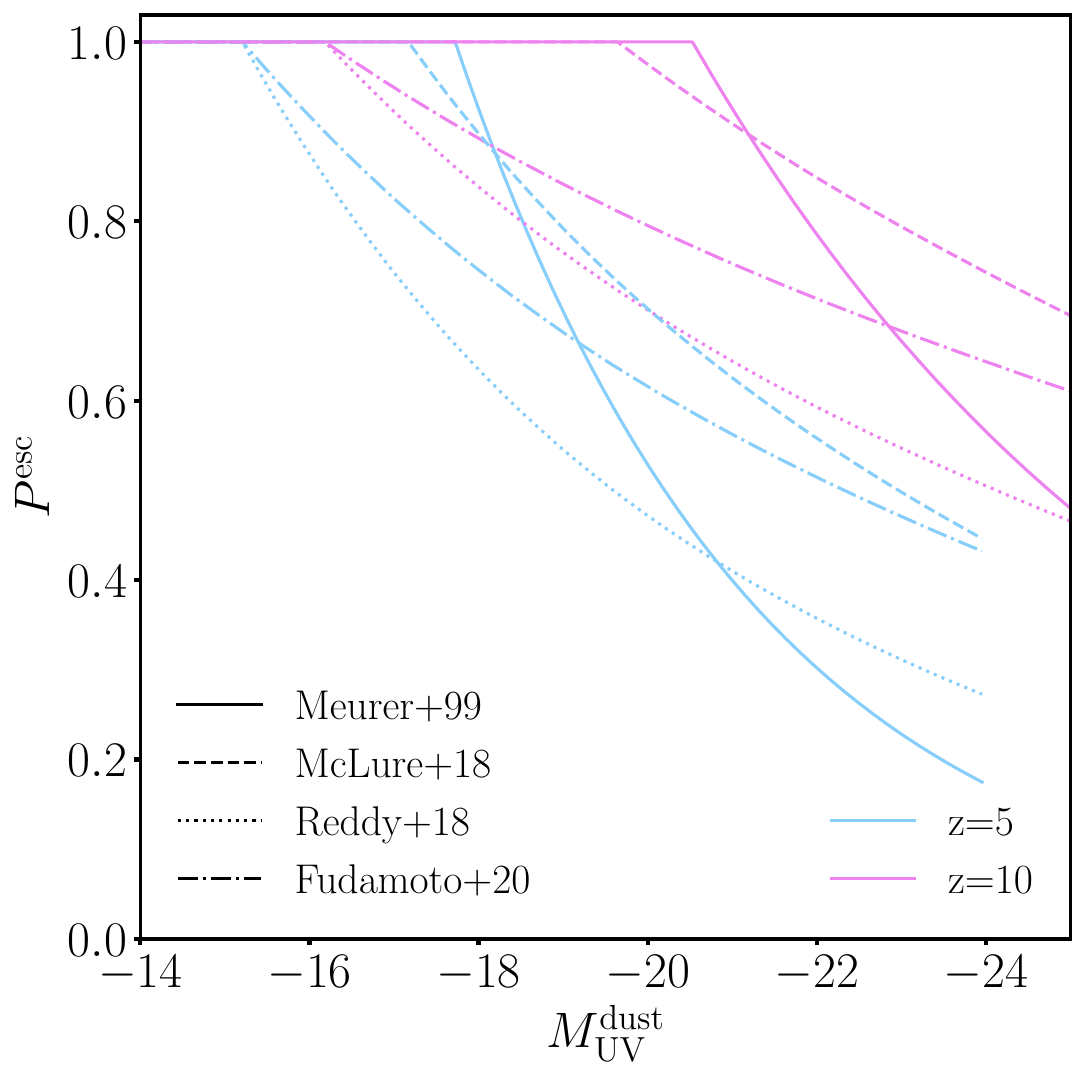}
    \caption{The effect of the assumed IRX-$\beta$ relation on the UV photon escape probability at $z=5$ and~10. The solid curves are calculated from \cite{Meurer1999} derived from local starburst galaxies. The dotted curves use an SMC-like extinction curve calibrated to galaxies at $z \sim 2$--3  \cite{Reddy2018}. The dashed curves use a ``grey attenuation'' curve calibrated to galaxies at $z \sim 2$--3 \cite{McLure2018}. The dash-dotted curves are estimated from \cite{Fudamoto2020} using the ALPINE survey at $z \sim 5$.}
    \label{fig:IRX}
\end{figure}

A key part of our framework is the empirical calibration of the UV attenuation. However, this quantity is not measured directly in the existing samples but is inferred from the UV spectral indices, based on a calibration to a sample at much lower redshifts. As noted in section~\ref{results}, recent work shows that this calibration may break down at high redshifts \cite{Fudamoto2020}. To further examine the effects of particular IRX-$\beta$ relations on our model, we show in Figure~\ref{fig:IRX} how different IRX-$\beta$ calibrations affect the UV photon escape probabilities. The solid lines show our fiducial model (from local starburst galaxies; \cite{Meurer1999}), the dashed and dotted lines show estimates from galaxies at $z\sim2-3$ \cite{McLure2018, Reddy2018} that adopt two types of extinction curves (the ``grey'' curve and the ``SMC-like'' curve, respectively), and the dash-dotted curves show an estimate from ALPINE galaxies at $z \sim 5$ \cite{Fudamoto2020}.

\begin{figure}
    \centering
	\includegraphics[width=0.5\textwidth]{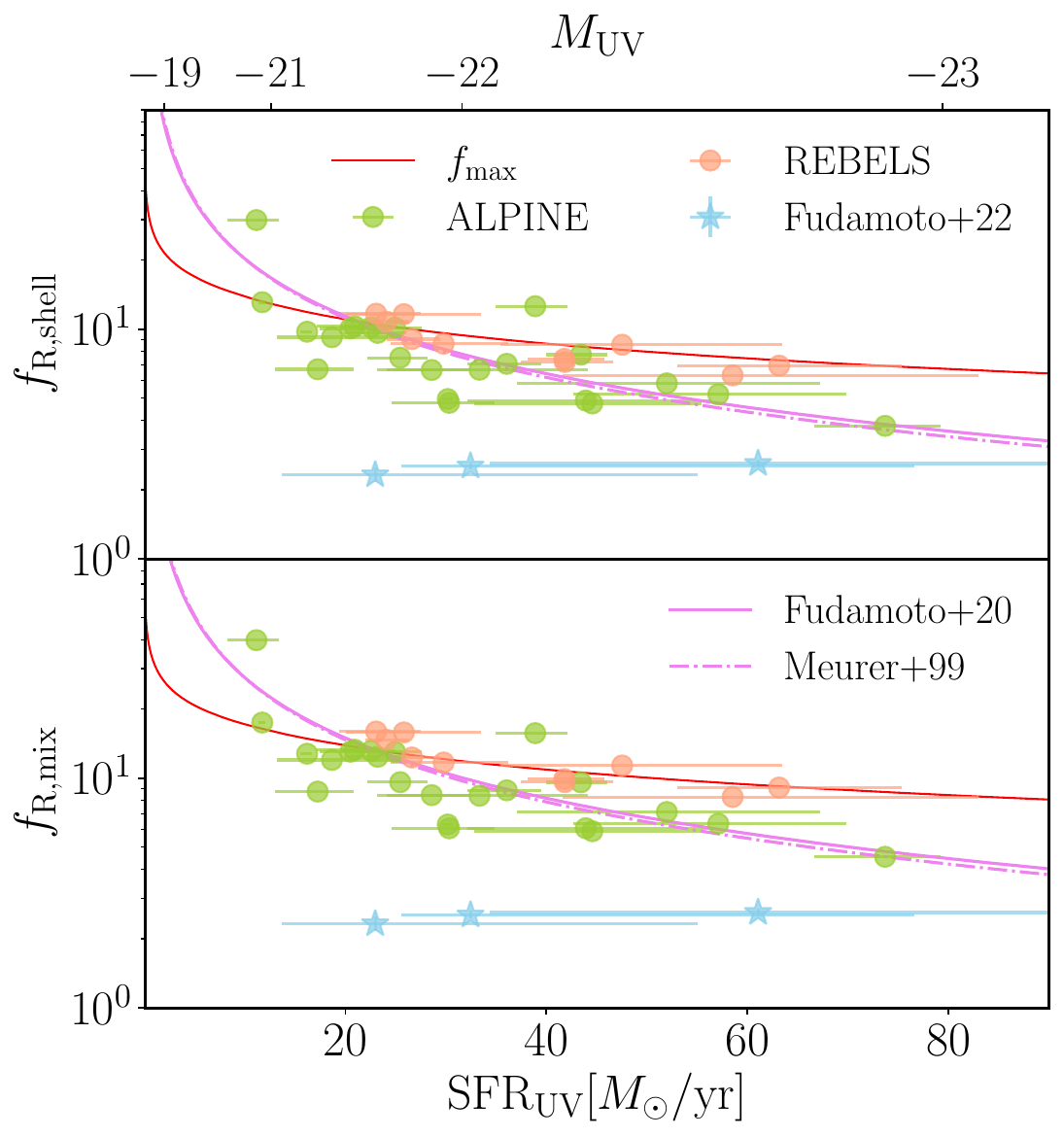}
    \caption{Estimated radius scaling parameters if we use \cite{Fudamoto2020} to infer the level of UV attenuation (c.f. the right panels of Fig.~\ref{fig:fmfr}, which used the relation from \cite{Meurer1999}, shown here by the dash-dotted curves).}
    \label{fig:Fud_IRX}
\end{figure}

Interestingly, our fiducial model finds the steepest dependence of attenuation on galaxy luminosity: all the models calibrated to higher redshift galaxies have gentler slopes, with the flattest resulting from calibration to ALPINE galaxies \cite{Fudamoto2020}. The estimates at $z \sim 2$--3 are in between these limits. The gentler slopes mean that small galaxies still have a modest amount of dust, though it is still a very small effect on the overall UV luminosity. We re-run the $f_R$ analysis using \cite{Fudamoto2020} and show the results in Figure~\ref{fig:Fud_IRX}. Note that the $f_M$ results are independent of the IRX-$\beta$ choice (because they are derived from the IR emission), so we only show the change in $f_R$. It is obvious that a gentler IRX-$\beta$ slope increases $f_R$ compared to our fiducial model, increasing the tension with the dust radii inferred from stacking \cite{Fudamoto2022}.

%%%%%%%%% FIGURE: Fudamoto fM, fR 

We note as well that the recent study of \cite{Cullen2023a} found a steep evolution of the UV slope at $z\gtrsim 9$ (Figure~\ref{fig:betafit}). Along the same lines, \cite{Reddy2018} has suggested the intrinsic UV slope could become bluer towards higher redshift. If so, this would compromise the extrapolation of any of these results to very high redshifts during the reionization era. A full understanding of the dust will therefore hinge on a better understanding of the intrinsic UV slopes of these stellar populations.

\subsection{Dust and Bursty Galaxies}
\label{bursts}

Another limitation of our model is the assumption of a one-to-one correspondence between halo mass and star formation rate. In reality, starbursts appear to be a crucial part of high-$z$ galaxy evolution \cite{Mason2023, Mirocha2023}. The ultimate cause of this burstiness is not clear, but one possibility is that the short dynamical times of high-$z$ galaxies allow star formation to ``overshoot'' the expectations from feedback regulation \cite{Faucher-Giguere2018, Furlanetto2022, Dekel2023}. The excess feedback energy from the runaway star formation can then shut down that process for a period by driving gas (and presumably dust) outside of the host halo, until the cycle begins anew. This process can occur in small galaxies \cite{Furlanetto2022} and large ones \cite{Dekel2023}. 

As a result, the mapping between star formation and stellar mass can have large fluctuations, which we have thus far ignored. In this section, we will estimate the effects of such scatter in a simple phenomenological way. We assume that the total dust mass is unaffected by burstiness, because it depends on the stellar mass in our model (which is integrated over all past episodes of star formation so is not as strongly affected by the current SFR). However, the instantaneous luminosity -- and hence the energy input into the dust phase -- does depend on the current SFR. 

For a toy model, we assume that the intrinsic luminosity of galaxies varies by a fixed amount of $\pm 1$ or $2$ magnitudes. The former is comparable to the scatter expected in supernova-delay induced burstiness \cite{Furlanetto2022}, while the latter is close to the required level to explain the bright end of the luminosity function at high redshifts \cite{Mason2023, Mirocha2023}. We then use this luminosity in equation~(\ref{eq:T_dust}) to compute the dust temperature, from which the dust emission follows. (Note that here we assume that the fraction of absorbed starlight does not change from the fiducial calculation; see below.) 

%%%%%%%%%% FIGURE: dust emission in bursts
\begin{figure}
    \centering
	\includegraphics[width=0.5\textwidth]{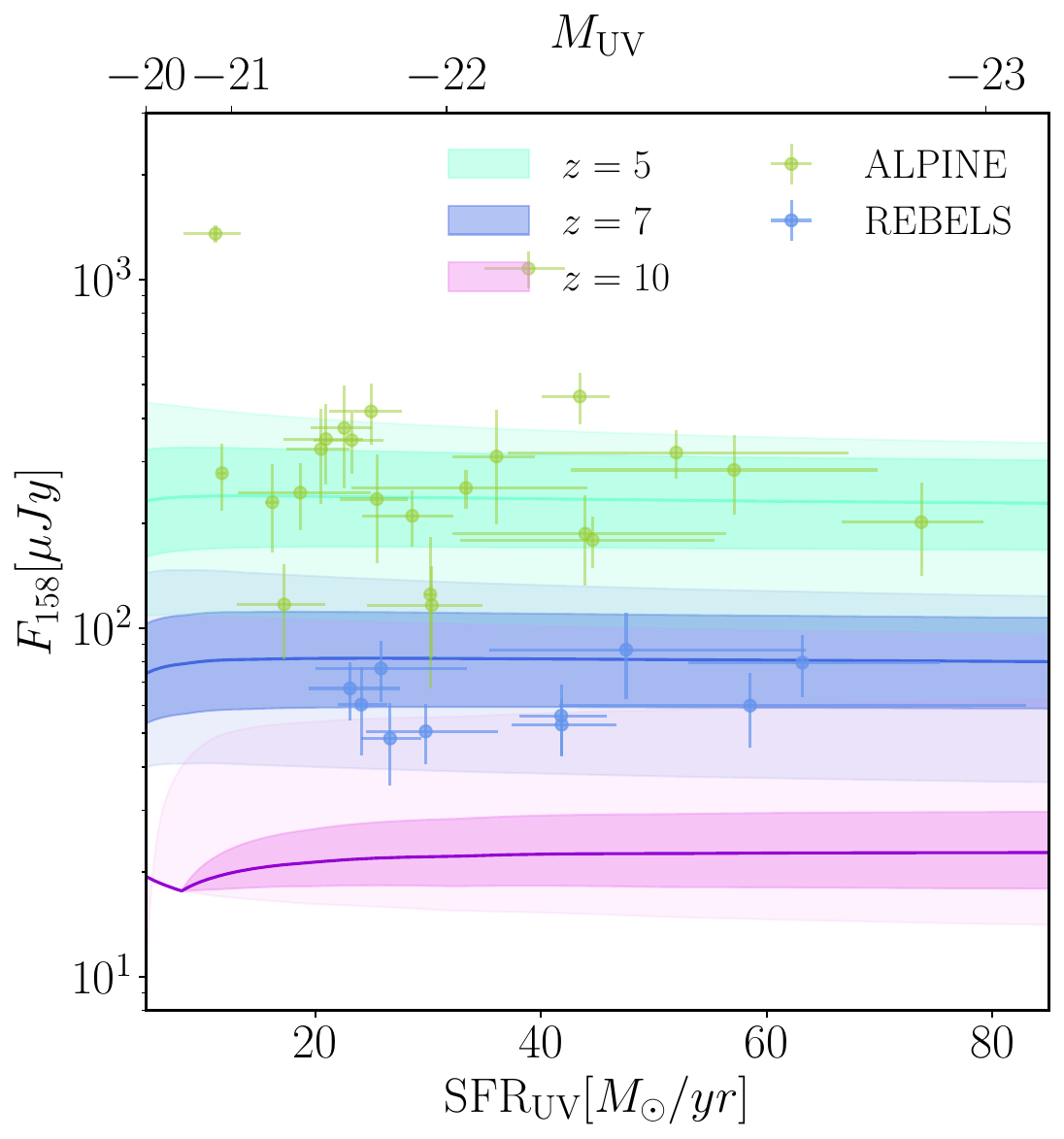}
    \caption{The burstiness-induced scatter in [\cii] continuum flux at various redshifts as a function of $\mathrm{SFR_{UV}}$. The solid curves are the average continuum flux. The shaded region highlights the scattering induced in these quantities at a given redshift with (1, 2) magnitudes. The darker shades are the 1 magnitude scatters, and shallower shades are the 2 magnitude scatters. The green circles are the ALPINE results from \cite{Bethermin2020}. The blue circles are the REBELS results from \cite{Ferrara2022}. }
    \label{fig:Bursty_Rabs}
\end{figure}

Figure~\ref{fig:Bursty_Rabs} shows the resulting spread in the observed flux for several redshifts. We find that a moderate amount of burstiness can easily explain the observed scatter in the REBELS sample. The ALPINE galaxies show a wider range, however. We note that scatter of $\pm 1$~mag causes a scatter in the dust temperatures of $\pm 5$~K as well. 

Here we have assumed for simplicity that the dust spatial distribution is unaffected by burstiness. If the dust distribution is primarily a function of the total stellar mass, then our UV-calibration procedure would imply that the escape fraction of UV photons also has scatter at each UV luminosity. Also, we note that it may actually help explain gaps in the dust morphology, as we appear to require (see also \cite{Inami2022, Fudamoto2022}). The increased radiation pressure during a starburst can drive dust away from the stars, segregating it from the UV emission, especially in small systems \cite{Tsuna2023, Ziparo2023}.

\subsection{Dust and the UVLF}

One of the key surprises of JWST's early surveys is the unexpected abundance of luminous systems at $z \gtrsim 10$. While burstiness provides one potential explanation, dust may contribute as well \cite{Ferrara2023, Ziparo2023}. The key idea is that the abundance of UV-luminous systems at $z \sim 7$ is suppressed by their relatively large amounts of dust, but that at higher redshifts the amount of dust decreases, partly because it may be ejected by strong feedback \cite{Ziparo2023}. Thus if we normalize our model parameters to match the low-$z$ data, the galaxy abundance will decrease less than expected at higher redshifts because the decreasing dust obscuration partially counteracts the decrease in the halo mass function. 

\begin{figure}
    \centering
	\includegraphics[width=0.5\textwidth]{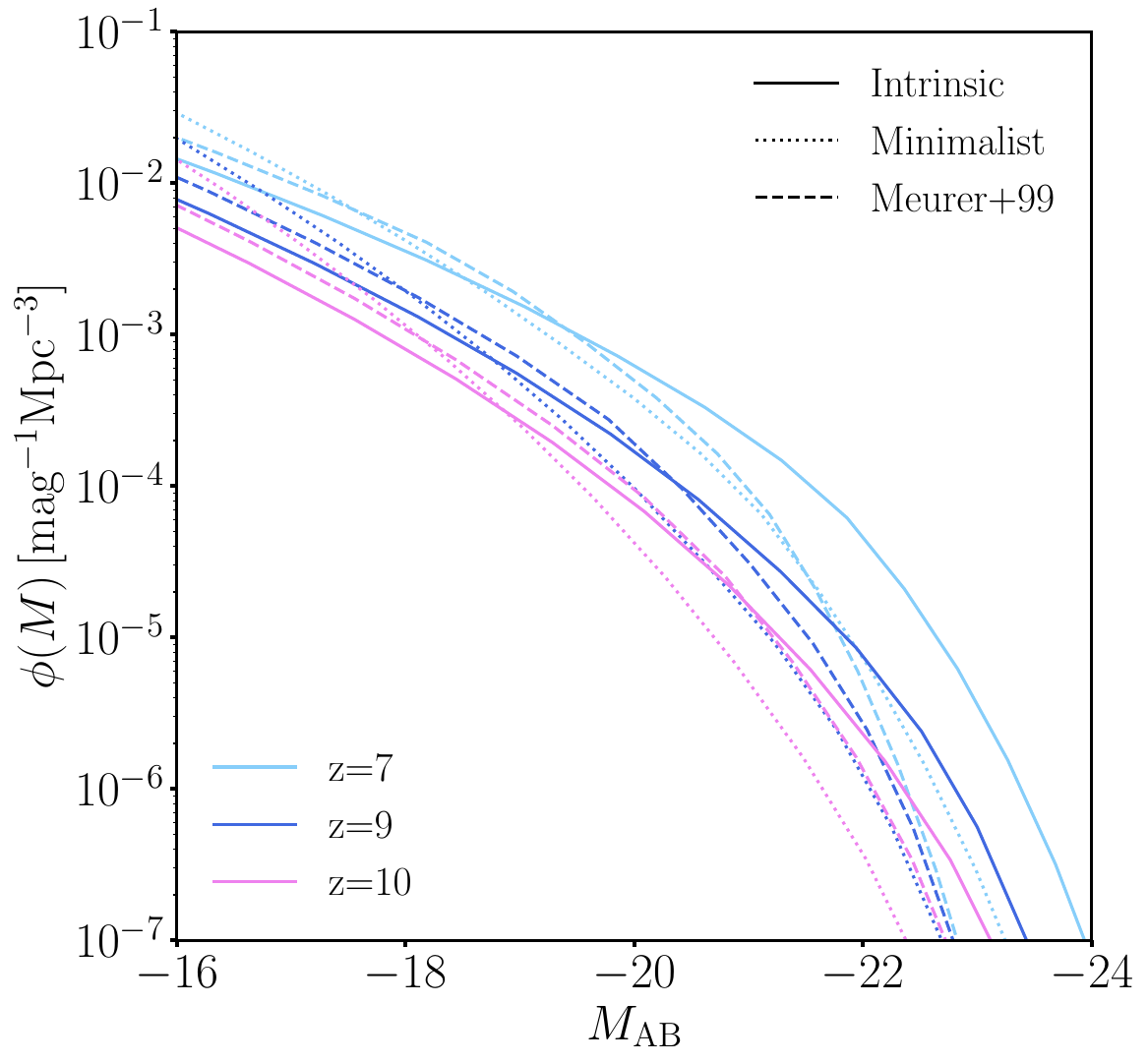}
    \caption{The effect of dust on the high-$z$ UVLF. The dotted curves show the UVLF computed from \cite{Furlanetto2017}, which ignores dust. The solid curves show the \emph{intrinsic} UVLF without dust-processing, assuming the IRX-$\beta$ relation from \cite{Meurer1999}. The dashed curves are LFs recovered from the \emph{intrinsic} by applying the \cite{Meurer1999}  IRX-$\beta$ relation. Note that the effects of dust obscuration are quite modest by $z \sim 10$.} 
    \label{fig:Bursty_Mason}
\end{figure}

We now consider this explanation in the context of our model. Figure~\ref{fig:Bursty_Mason} shows two sets of luminosity functions: one in which we ignore dust entirely (and is thus equivalent to the minimalist model of \cite{Furlanetto2017} as the ``observed'' LFs; dotted curves) and one in which we include our dust model with the IRX-$\beta$ relation from \cite{Meurer1999}. Here we have obtained the intrinsic LF at $z \sim 7$ by correcting to the total SFR using the \cite{Meurer1999} relation. The solid curves show this intrinsic LF from $z=7$--10; note that this is \emph{not} the observable UVLF, because it includes star formation hidden by dust. The dashed curves then show the expected LF after dust attenuation is included. 

We see that, as expected, dust attenuation is significant at the bright end at $z \sim 7$, but by $z \sim 10$ it has a much smaller impact. This conclusion matches the current consensus that early galaxies at $z\sim8-10$ are not expected to contain abundant amounts of dust \cite{Ferrara2023, Mason2023, Ziparo2023}. Importantly, the dashed and dotted curves are near each other at all redshifts -- that is, the overall redshift evolution is comparable when we ignore and include dust. We do not find that it can account entirely for the overabundance of bright galaxies at high redshifts, although it does help to some extent. This emphasizes how dust evolution can complement other explanations for the excess of sources, and it may even be physically associated with burstiness (as described in the previous subsection). 

\subsection{The Fraction of Total Obscured SFR}
\label{SFRD}

So far, we have focused on dust in individual galaxies. We now shift to the cumulative effect of dust across the entire galaxy population by considering the star formation rate density (SFRD), which is given by 
\begin{equation}
    \rho_{\rm tot}(z) = \int dm\, \dot{M}_{\rm \star, tot}(m, z)n(m, z).
\end{equation}
where the integration variable $m$ is the halo mass and $n$ is the halo comoving number density in the mass range $(m, m+dm)$ at redshift $z$. We calculate both the total SFRD and the obscured SFRD $\rho_{\rm obs}$, so that the obscured fraction is $f_{\rm obsc}=\rho_{\rm obsc}/ \rho_{\rm tot}$. We calculate this quantity for the entire halo population that is able to form stars. We suppose that the lowest mass limit for a halo to form stars is given by its virial temperature $T_{\rm vir} \gtrsim 10^{4}{\rm K}$. (We take the virial temperature formulation from \cite{Barkana2001}'s equation (26), which roughly gives a halo mass limit of $10^8M_\odot$.) We also calculate the SFRD for a subset of massive halos (with a minimum at $10^{11} \ M_\odot$). The latter corresponds roughly to $M_{\rm UV}\lesssim-19$, so it better reflects the fraction that is directly observable.

\begin{figure}
    \centering
	\includegraphics[width=0.5\textwidth]{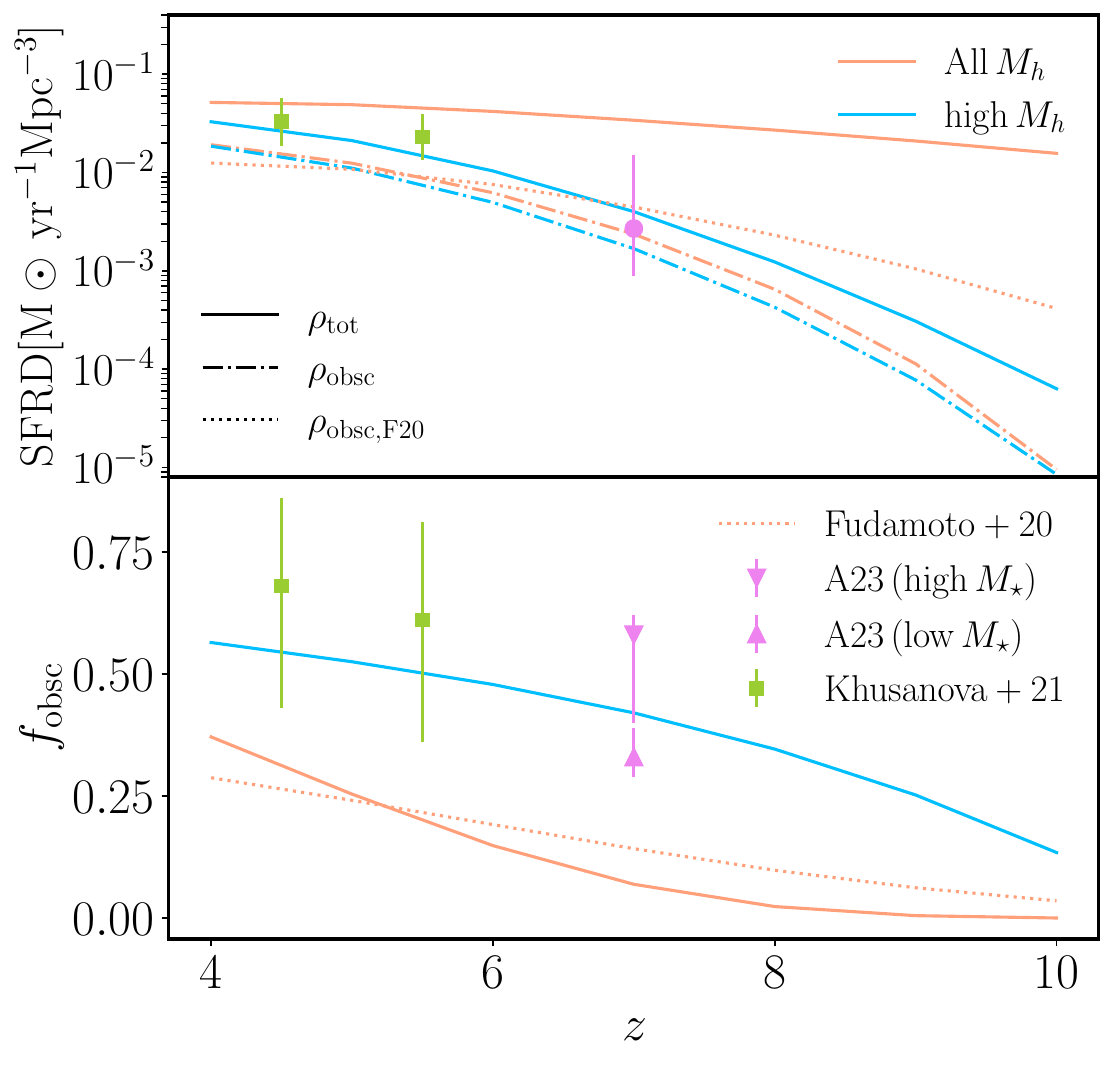}
    \caption{Effect of dust obscuration on the global SFRD as a function of redshift. In the upper panel, the orange curves show the SFRD including all halos able to form stars (roughly $10^{8}- 10^{13} \ M_\odot$ in our model), while blue curves only include massive halos ($10^{11}- 10^{13}M_\odot$, which corresponds to $M_{\rm UV}\lesssim-19$ at $z\approx5$ so are more easily observable). The solid curves are the intrinsic SFRD, the dash-dotted curves are the dust-obscured SFRD from \cite{Meurer1999}, and the dotted curves are the obscured SFRD computed from \cite{Fudamoto2020}'s IRX-$\beta$ relation for all halo masses. The circle symbol denotes estimated SFRD from the REBELS sample \cite{Algera2023}. In the lower panel, we show $f_{\rm obsc}$ directly. The orange curve and the blue curve are the $f_{\rm obsc}$ calculated from all halos and massive halos, respectively. The dotted orange curve is also $f_{\rm obsc}$ calculated from all halos, but from \cite{Fudamoto2020}'s IRX-$\beta$ relation. The purple symbols are \cite{Algera2023}'s stacking analysis of the obscured fraction in observed galaxies, with the downward triangle representing their higher stellar mass stack and the upward symbol representing their lower stellar mass stack. The green squares are the ALPINE stacking analysis from \cite{Khusanova2021}.}
    \label{fig:fobs}
\end{figure}

Figure~\ref{fig:fobs} shows the results, with the SFRD in the top panel and the obscured fraction in the bottom panel. We see that at the low end of the redshift range a substantial fraction of the star formation is hidden by dust, especially in the massive galaxies. But our model predicts that by $z \sim 10$ obscuration is a modest effect overall. In general, obscuration is minimal in faint galaxies across this redshift interval, so if they dominate the photon budget for, e.g., reionization \cite{Simmonds2023}, dust plays only a relatively minor role. We note that this remains true even using the IRX-$\beta$ relation from \cite{Fudamoto2020}, which has a gentler dependence of the dust mass on luminosity. Even though the faint galaxies have more dust according to their prescription, it is not enough to increase the overall SFRD significantly.

We also compare our results to estimates based on observed sources from \cite{Khusanova2021} and \cite{Algera2023}.  They are quite consistent with our model, at least for massive galaxies, but we emphasize that the (dominant) low-mass galaxies can differ substantially, especially at high redshifts. We also note that all of these results (including our own) are based on UV-selected sources. \cite{Williams2023} has suggested that the presence of UV-faint but massive galaxies, undetected by HST, may contribute significantly to the obscured cosmic SFRD at $z\gtrsim4$, resulting in an underestimation of that quantity. Spectroscopic surveys across diverse galaxy populations will ultimately be needed to better constrain the obscured contribution.

\section{Conclusions}
\label{conclusions}

We have presented a semi-analytic model of dust in high-$z$ galaxies, aiming to constrain its parameters by simultaneously matching the observed UV and IR properties of galaxies during reionization. Our dust model pairs a simple galaxy evolution model \cite{Furlanetto2017} with a streamlined treatment of the dust distribution \cite{Imara2018, Mirocha2020} to generate self-consistent predictions for the UV attenuation and far-IR continuum emission. The key input parameters are the dust mass and radial distribution. We constrain these quantities with the UV attenuation inferred from the IRX-$\beta$ relation \cite{Meurer1999} and from ALMA measurements of the far-IR continuum flux. We find:

\emph{(i)} The dust attenuation increases with galaxy mass and decreases with redshift, as expected \cite{Ferrara2022, Bowler2023}. In our model, very few galaxies have significant attenuation beyond $z \sim 10$. The observed galaxies at $z \lesssim 10$, which are of course relatively bright, suffer from non-negligible attenuation. Nevertheless, our model predicts that these modest extinctions are far less than one expects from normal dust production models, if the dust and stars are confined to rotation-supported disks. 

\emph{(ii)} By simultaneously calibrating to the UV attenuation and IR emission, our model requires that the total dust mass be close to the amount produced by normal stellar populations but that the dust be significantly more extended than a rotationally-supported disk. This conclusion holds regardless of whether we assume the dust to be distributed in a shell surrounding the galaxy or mixed throughout the stellar population. Interestingly, we find that dust may be better retained (albeit with a larger relative extent) in the smallest galaxies, as previously suggested by \cite{Dayal2022, Topping2022, Palla2023} but in contrast to expectations from standard feedback models.

\emph{(iii)} However, the inferred radii of the dust distribution are larger than those measured by \cite{Fudamoto2022} by a factor of a few. One way to reconcile these results -- while maintaining the relatively large dust luminosities -- is if the dust has a more complicated distribution, with the dust and young stars segregated from each other. In that case, if the dust is distributed on the spatial scale suggested by \cite{Fudamoto2022}, we find typical dust covering fractions  $f_{\rm cov} \sim 0.2$--$0.7$ for massive galaxies (similar to many ALMA-derived models; e.g. \cite{Khusanova2021, Dayal2022, Ferrara2022, Algera2023}), while faint galaxies have $f_{\rm cov} \approx 0$. This could result from feedback ejecting the dust locally from star-forming regions but not from the halo.

Even in our anisotropic model, the dust is still a few times more spread out than the UV emission (though it is comparable to the extent of the CII emission; \cite{Pozzi2024}). In the context of our model, this occurs because feedback drives dust out of its formation region. This presents a challenge, as hot winds should evaporate dust. Nevertheless, dust can be found in some winds \cite{Bland-Hawthorn2003, Rupke2013, Triani2021, Katsioli2023}, which as recently been examined from a theoretical perspective by \cite{Richie2024}. It will be important to understand these mechanisms as we learn more about dust in high-$z$ galaxies.

Our results have implications not only for dust evolution but potentially for other aspects of early galaxy formation. For example, reionization requires a substantial fraction of ionizing photons to escape from early galaxies. If indeed feedback has cleared out channels in the dust distribution, it may also have cleared out neutral gas -- allowing these ionizing photons to escape more easily. Additionally, the large dust luminosities suggest that most of the dust is retained in the galaxy's halo, which means feedback may not be effective at expelling metals to large distances (c.f. \cite{Yamaguchi2023}). 

However, there are several caveats to our model. Most importantly, our galaxy formation model associates halos to star formation histories (and dust properties) on a one-to-one basis, such that the SFR is a monotonically increasing function of halo mass. Real halos have a substantial scatter in their SFRs, which makes a comparison with observations challenging. Using a simple treatment of burstiness, we found that it can explain the scatter in the observations, but it is not clear if it is responsible for the apparent decrease in the retention fraction of dust as a function of SFR. Additionally, we have assumed that the IRX--$\beta$ relation from low-$z$ galaxies \cite{Meurer1999} applies to high-$z$ sources. We have checked that our qualitative conclusions hold when other relations, based on higher-redshift sources \cite{McLure2018, Reddy2018, Fudamoto2020}, are used, but the quantitative results are sensitive to this relation. Finally, while our results are suggestive of a complex dust morphology, we do not attempt to model it or explore the physical mechanism behind it. Future observations will no doubt shed light on all of these issues.

\appendix

\section{The UV Slope--Luminosity Relation in Early JWST Surveys}
\label{appendix}

%%%%%%%%%%%% FIGURE: UV slopes - appendix
\begin{figure*}
	\includegraphics[width=\textwidth]{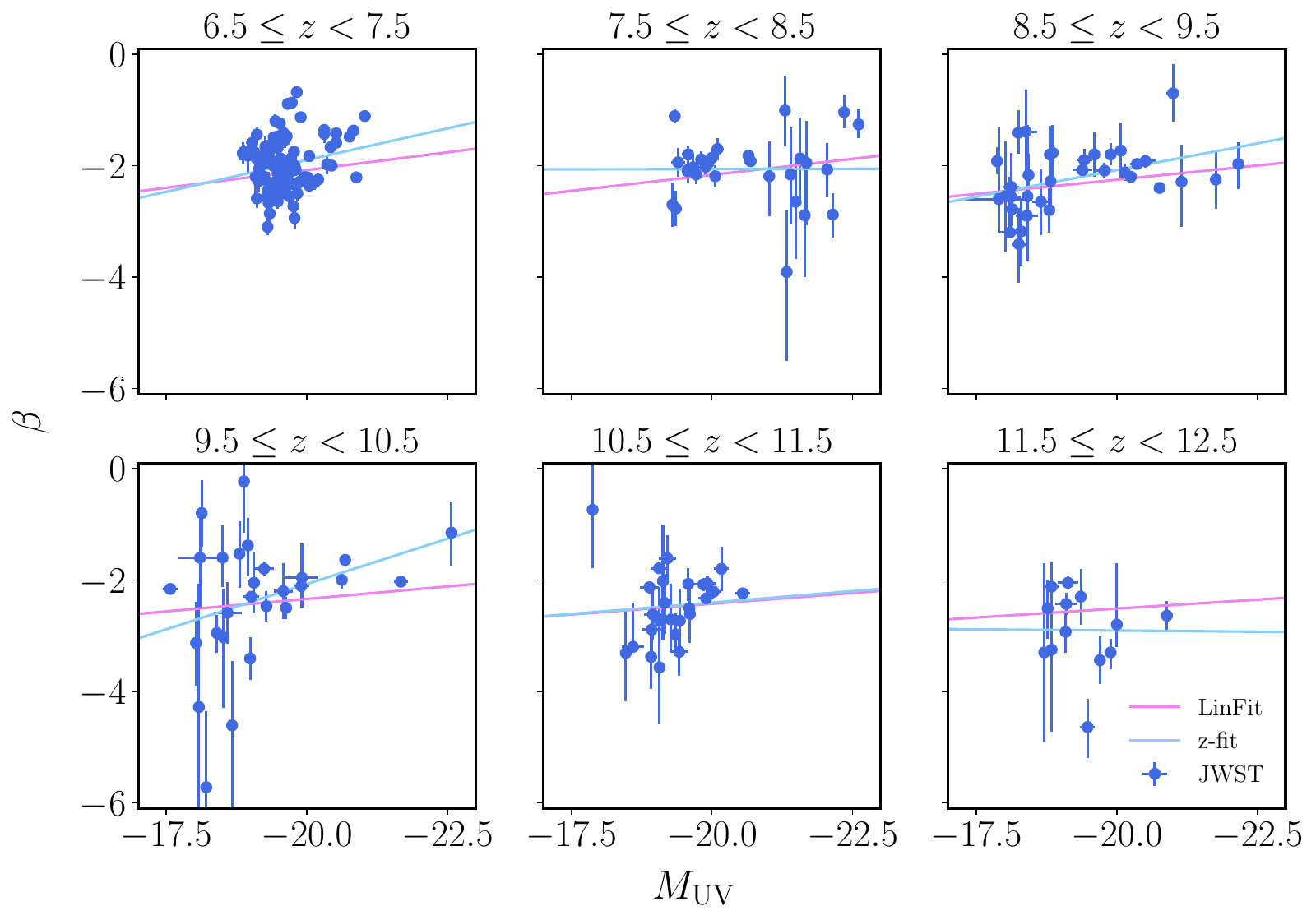}
    \caption{The $\beta$-$M_\mathrm{UV}$ relation at redshifts $6.5\leq z \leq 12.5$ separated into six redshift bins and compared to measurements from the literature (blue points; see text for a full list of sources). The purple curve represents the best fit relation to the \cite{Bouwens2014} and \cite{Topping2023a} measurements (as used in the main text), while the light blue curves show the best fit to data displayed here in each redshift interval. }
    \label{fig:betaz}
\end{figure*}

In the main text, we based our UV slope fit on two large homogeneous samples \cite{Bouwens2014, Topping2023a}, but other measurements are available in the literature. To check for any obvious biases, here we compare our fits to other samples, which we have assembled from ERO and ERS NIRCam imaging and ground-based near-infrared imaging of the COSMOS field \cite{Cullen2023, Donnan2023}, with additional pointings in the Hubble ultra deep field (HUDF) field and GOODS-South field \cite{Donnan2023a}, the UNCOVER survey \cite{Atek2023}, the JEMS survey \cite{Bouwens2023}, the NGDEEP field \cite{Austin2023}, and the CEERS survey \cite{Endsley2023, Whitler2023}. Figure~\ref{fig:betaz} shows that our fit is largely consistent with these other measurements, although the scatter is large, even at redshifts beyond the $z \sim 10$ limit used in this work.

\acknowledgments
This work was supported by NASA through award 80NSSC22K0818 and by the National Science Foundation through awards AST-1812458 and AST-2205900. We thank Hiddo S. B. Algera, Yana Khusanova, and Michael Topping for providing their data, as well as Sahil Hegde and Adam Trapp for useful discussions. RJZ acknowledges the fellowship support from the James and Meredith Henry Endowment administered through the UCLA Undergraduate Research Scholars Program (URSP) under the UCLA Undergraduate Research Center - Sciences. RJZ thanks Meadowridge School for hospitality while much of this work was completed, along with Dale C. Stevenson for helpful conversations. This work has made extensive use of NASA's Astrophysics Data System (\href{http://ui.adsabs.harvard.edu/}{http://ui.adsabs.harvard.edu/}) and the arXiv e-Print service (\href{http://arxiv.org}{http://arxiv.org}), as well as the following softwares: \textsc{matplotlib} \cite{Matplotlib}, \textsc{numpy} \cite{Numpy}, \textsc{Pandas} \cite{Pandas}, and \textsc{scipy} \cite{Scipy}.

The authors wish to acknowledge our presence on the lands in which this work is produced. The authors at UCLA Department of Physics \& Astronomy acknowledge our presence on the traditional, ancestral, and unceded territory of the Gabrielino/Tongva peoples. Much of this work is also completed at Meadowridge School, located on the ancestral and unceded territories of the Katzie, Kwantlen, and Coast Salish Peoples. We value the opportunity to learn, live, and share research and educational experiences on these traditional lands.

\bibliographystyle{JHEP}
\bibliography{citation}

\providecommand{\href}[2]{#2}\begingroup\raggedright\begin{thebibliography}{100}

\bibitem{Atek2023}
H.~{Atek}, I.~{Chemerynska}, B.~{Wang}, L.~{Furtak}, A.~{Weibel}, P.~{Oesch} et~al., \emph{{JWST UNCOVER: Discovery of $z>9$ Galaxy Candidates Behind the Lensing Cluster Abell 2744}}, \href{https://doi.org/10.48550/arXiv.2305.01793}{\emph{arXiv e-prints} (2023) arXiv:2305.01793} [\href{https://arxiv.org/abs/2305.01793}{{\ttfamily 2305.01793}}].

\bibitem{Austin2023}
D.~{Austin}, N.J.~{Adams}, C.J.~{Conselice}, T.~{Harvey}, K.~{Ormerod}, J.~{Trussler} et~al., \emph{{A Large Population of Faint 8<z<16 Galaxies Found in the First JWST NIRCam Observations of the NGDEEP Survey}}, \href{https://doi.org/10.48550/arXiv.2302.04270}{\emph{arXiv e-prints} (2023) arXiv:2302.04270} [\href{https://arxiv.org/abs/2302.04270}{{\ttfamily 2302.04270}}].

\bibitem{Bouwens2023}
R.J.~{Bouwens}, M.~{Stefanon}, G.~{Brammer}, P.A.~{Oesch}, T.~{Herard-Demanche}, G.D.~{Illingworth} et~al., \emph{{Evolution of the UV LF from z 15 to z 8 using new JWST NIRCam medium-band observations over the HUDF/XDF}}, \href{https://doi.org/10.1093/MNRAS,/stad1145}{\emph{MNRAS,} {\bfseries 523} (2023) 1036} [\href{https://arxiv.org/abs/2211.02607}{{\ttfamily 2211.02607}}].

\bibitem{Endsley2023}
R.~{Endsley}, D.P.~{Stark}, L.~{Whitler}, M.W.~{Topping}, Z.~{Chen}, A.~{Plat} et~al., \emph{{A JWST/NIRCam Study of key contributors to reionization: The star-forming and ionizing properties of UV-faint z 7 - 8 galaxies}}, \href{https://doi.org/10.1093/MNRAS,/stad1919}{\emph{MNRAS,} (2023) } [\href{https://arxiv.org/abs/2208.14999}{{\ttfamily 2208.14999}}].

\bibitem{Tacchella2023}
S.~{Tacchella}, B.D.~{Johnson}, B.E.~{Robertson}, S.~{Carniani}, F.~{D'Eugenio}, N.~{Kumari} et~al., \emph{{JWST NIRCam + NIRSpec: interstellar medium and stellar populations of young galaxies with rising star formation and evolving gas reservoirs}}, \href{https://doi.org/10.1093/MNRAS,/stad1408}{\emph{MNRAS,} {\bfseries 522} (2023) 6236} [\href{https://arxiv.org/abs/2208.03281}{{\ttfamily 2208.03281}}].

\bibitem{Topping2023a}
M.W.~{Topping}, D.P.~{Stark}, R.~{Endsley}, L.~{Whitler}, K.~{Hainline}, B.D.~{Johnson} et~al., \emph{{The UV Continuum Slopes of Early Star-Forming Galaxies in JADES}}, {\emph{arXiv e-prints} (2023) arXiv:2307.08835} [\href{https://arxiv.org/abs/2307.08835}{{\ttfamily 2307.08835}}].

\bibitem{Whitler2023}
L.~{Whitler}, R.~{Endsley}, D.P.~{Stark}, M.~{Topping}, Z.~{Chen} and S.~{Charlot}, \emph{{On the ages of bright galaxies 500 Myr after the big bang: insights into star formation activity at z {\ensuremath{\gtrsim}} 15 with JWST}}, \href{https://doi.org/10.1093/MNRAS,/stac3535}{\emph{MNRAS,} {\bfseries 519} (2023) 157} [\href{https://arxiv.org/abs/2208.01599}{{\ttfamily 2208.01599}}].

\bibitem{Mason2023}
C.A.~{Mason}, M.~{Trenti} and T.~{Treu}, \emph{{The brightest galaxies at cosmic dawn}}, \href{https://doi.org/10.1093/MNRAS,/stad035}{\emph{MNRAS,} {\bfseries 521} (2023) 497} [\href{https://arxiv.org/abs/2207.14808}{{\ttfamily 2207.14808}}].

\bibitem{Mirocha2023}
J.~{Mirocha} and S.R.~{Furlanetto}, \emph{{Balancing the efficiency and stochasticity of star formation with dust extinction in z {\ensuremath{\gtrsim}} 10 galaxies observed by JWST}}, \href{https://doi.org/10.1093/MNRAS,/stac3578}{\emph{MNRAS,} {\bfseries 519} (2023) 843} [\href{https://arxiv.org/abs/2208.12826}{{\ttfamily 2208.12826}}].

\bibitem{Ferrara2023}
A.~{Ferrara}, A.~{Pallottini} and P.~{Dayal}, \emph{{On the stunning abundance of super-early, luminous galaxies revealed by JWST}}, \href{https://doi.org/10.1093/MNRAS,/stad1095}{\emph{MNRAS,} {\bfseries 522} (2023) 3986} [\href{https://arxiv.org/abs/2208.00720}{{\ttfamily 2208.00720}}].

\bibitem{Ziparo2023}
F.~{Ziparo}, A.~{Ferrara}, L.~{Sommovigo} and M.~{Kohandel}, \emph{{Blue monsters. Why are JWST super-early, massive galaxies so blue?}}, \href{https://doi.org/10.1093/MNRAS,/stad125}{\emph{MNRAS,} {\bfseries 520} (2023) 2445} [\href{https://arxiv.org/abs/2209.06840}{{\ttfamily 2209.06840}}].

\bibitem{Inami2022}
H.~{Inami}, H.S.B.~{Algera}, S.~{Schouws}, L.~{Sommovigo}, R.~{Bouwens}, R.~{Smit} et~al., \emph{{The ALMA REBELS Survey: dust continuum detections at z > 6.5}}, \href{https://doi.org/10.1093/MNRAS,/stac1779}{\emph{MNRAS,} {\bfseries 515} (2022) 3126} [\href{https://arxiv.org/abs/2203.15136}{{\ttfamily 2203.15136}}].

\bibitem{Calzetti2001}
D.~{Calzetti}, \emph{{The Dust Opacity of Star-forming Galaxies}}, \href{https://doi.org/10.1086/324269}{\emph{PASP} {\bfseries 113} (2001) 1449} [\href{https://arxiv.org/abs/astro-ph/0109035}{{\ttfamily astro-ph/0109035}}].

\bibitem{Weingartner2001}
J.C.~{Weingartner} and B.T.~{Draine}, \emph{{Dust Grain-Size Distributions and Extinction in the Milky Way, Large Magellanic Cloud, and Small Magellanic Cloud}}, \href{https://doi.org/10.1086/318651}{\emph{ApJ,} {\bfseries 548} (2001) 296} [\href{https://arxiv.org/abs/astro-ph/0008146}{{\ttfamily astro-ph/0008146}}].

\bibitem{Schouws2022}
S.~{Schouws}, M.~{Stefanon}, R.~{Bouwens}, R.~{Smit}, J.~{Hodge}, I.~{Labb{\'e}} et~al., \emph{{Significant Dust-obscured Star Formation in Luminous Lyman-break Galaxies at z 7-8}}, \href{https://doi.org/10.3847/1538-4357/ac4605}{\emph{ApJ,} {\bfseries 928} (2022) 31} [\href{https://arxiv.org/abs/2105.12133}{{\ttfamily 2105.12133}}].

\bibitem{Meurer1999}
G.R.~{Meurer}, T.M.~{Heckman} and D.~{Calzetti}, \emph{{Dust Absorption and the Ultraviolet Luminosity Density at z \raisebox{-0.5ex}\textasciitilde 3 as Calibrated by Local Starburst Galaxies}}, \href{https://doi.org/10.1086/307523}{\emph{ApJ,} {\bfseries 521} (1999) 64} [\href{https://arxiv.org/abs/astro-ph/9903054}{{\ttfamily astro-ph/9903054}}].

\bibitem{Finkelstein2012}
S.L.~{Finkelstein}, C.~{Papovich}, B.~{Salmon}, K.~{Finlator}, M.~{Dickinson}, H.C.~{Ferguson} et~al., \emph{{Candels: The Evolution of Galaxy Rest-frame Ultraviolet Colors from z = 8 to 4}}, \href{https://doi.org/10.1088/0004-637X/756/2/164}{\emph{ApJ,} {\bfseries 756} (2012) 164} [\href{https://arxiv.org/abs/1110.3785}{{\ttfamily 1110.3785}}].

\bibitem{Casey2014}
C.M.~{Casey}, N.Z.~{Scoville}, D.B.~{Sanders}, N.~{Lee}, A.~{Cooray}, S.L.~{Finkelstein} et~al., \emph{{Are Dusty Galaxies Blue? Insights on UV Attenuation from Dust-selected Galaxies}}, \href{https://doi.org/10.1088/0004-637X/796/2/95}{\emph{ApJ,} {\bfseries 796} (2014) 95} [\href{https://arxiv.org/abs/1410.0702}{{\ttfamily 1410.0702}}].

\bibitem{McLure2018}
R.J.~{McLure}, J.S.~{Dunlop}, F.~{Cullen}, N.~{Bourne}, P.N.~{Best}, S.~{Khochfar} et~al., \emph{{Dust attenuation in 2 < z < 3 star-forming galaxies from deep ALMA observations of the Hubble Ultra Deep Field}}, \href{https://doi.org/10.1093/MNRAS,/sty522}{\emph{MNRAS,} {\bfseries 476} (2018) 3991} [\href{https://arxiv.org/abs/1709.06102}{{\ttfamily 1709.06102}}].

\bibitem{Reddy2018}
N.A.~{Reddy}, P.A.~{Oesch}, R.J.~{Bouwens}, M.~{Montes}, G.D.~{Illingworth}, C.C.~{Steidel} et~al., \emph{{The HDUV Survey: A Revised Assessment of the Relationship between UV Slope and Dust Attenuation for High-redshift Galaxies}}, \href{https://doi.org/10.3847/1538-4357/aaa3e7}{\emph{ApJ,} {\bfseries 853} (2018) 56} [\href{https://arxiv.org/abs/1705.09302}{{\ttfamily 1705.09302}}].

\bibitem{Fudamoto2020}
Y.~{Fudamoto}, P.A.~{Oesch}, A.~{Faisst}, M.~{B{\'e}thermin}, M.~{Ginolfi}, Y.~{Khusanova} et~al., \emph{{The ALPINE-ALMA [CII] survey. Dust attenuation properties and obscured star formation at z {\ensuremath{\sim}} 4.4-5.8}}, \href{https://doi.org/10.1051/0004-6361/202038163}{\emph{A\&A} {\bfseries 643} (2020) A4} [\href{https://arxiv.org/abs/2004.10760}{{\ttfamily 2004.10760}}].

\bibitem{Bowler2023}
R.A.A.~{Bowler}, H.~{Inami}, L.~{Sommovigo}, R.~{Smit}, H.S.B.~{Algera}, M.~{Aravena} et~al., \emph{{The ALMA REBELS survey: obscured star formation in massive Lyman-break galaxies at z = 4-8 revealed by the IRX-$\beta$ and $M_{\star}$ relations}}, {\emph{arXiv e-prints} (2023) arXiv:2309.17386} [\href{https://arxiv.org/abs/2309.17386}{{\ttfamily 2309.17386}}].

\bibitem{Khusanova2021}
Y.~{Khusanova}, M.~{Bethermin}, O.~{Le F{\`e}vre}, P.~{Capak}, A.L.~{Faisst}, D.~{Schaerer} et~al., \emph{{The ALPINE-ALMA [CII] survey. Obscured star formation rate density and main sequence of star-forming galaxies at z > 4}}, \href{https://doi.org/10.1051/0004-6361/202038944}{\emph{A\&A} {\bfseries 649} (2021) A152} [\href{https://arxiv.org/abs/2007.08384}{{\ttfamily 2007.08384}}].

\bibitem{Sommovigo2021}
L.~{Sommovigo}, A.~{Ferrara}, S.~{Carniani}, A.~{Zanella}, A.~{Pallottini}, S.~{Gallerani} et~al., \emph{{Dust temperature in ALMA [C II]-detected high-z galaxies}}, \href{https://doi.org/10.1093/MNRAS,/stab720}{\emph{MNRAS,} {\bfseries 503} (2021) 4878} [\href{https://arxiv.org/abs/2102.08950}{{\ttfamily 2102.08950}}].

\bibitem{Dayal2022}
P.~{Dayal}, A.~{Ferrara}, L.~{Sommovigo}, R.~{Bouwens}, P.A.~{Oesch}, R.~{Smit} et~al., \emph{{The ALMA REBELS survey: the dust content of z 7 Lyman break galaxies}}, \href{https://doi.org/10.1093/MNRAS,/stac537}{\emph{MNRAS,} {\bfseries 512} (2022) 989} [\href{https://arxiv.org/abs/2202.11118}{{\ttfamily 2202.11118}}].

\bibitem{Ferrara2022}
A.~{Ferrara}, L.~{Sommovigo}, P.~{Dayal}, A.~{Pallottini}, R.J.~{Bouwens}, V.~{Gonzalez} et~al., \emph{{The ALMA REBELS Survey. Epoch of Reionization giants: Properties of dusty galaxies at z {\ensuremath{\approx}} 7}}, \href{https://doi.org/10.1093/MNRAS,/stac460}{\emph{MNRAS,} {\bfseries 512} (2022) 58} [\href{https://arxiv.org/abs/2202.07666}{{\ttfamily 2202.07666}}].

\bibitem{Fudamoto2022}
Y.~{Fudamoto}, R.~{Smit}, R.A.A.~{Bowler}, P.A.~{Oesch}, R.~{Bouwens}, M.~{Stefanon} et~al., \emph{{The ALMA REBELS Survey: Average [C II] 158 {\ensuremath{\mu}}m Sizes of Star-forming Galaxies from z 7 to z 4}}, \href{https://doi.org/10.3847/1538-4357/ac7a47}{\emph{ApJ,} {\bfseries 934} (2022) 144} [\href{https://arxiv.org/abs/2206.01886}{{\ttfamily 2206.01886}}].

\bibitem{Algera2023}
H.S.B.~{Algera}, H.~{Inami}, P.A.~{Oesch}, L.~{Sommovigo}, R.J.~{Bouwens}, M.W.~{Topping} et~al., \emph{{The ALMA REBELS survey: the dust-obscured cosmic star formation rate density at redshift 7}}, \href{https://doi.org/10.1093/MNRAS,/stac3195}{\emph{MNRAS,} {\bfseries 518} (2023) 6142} [\href{https://arxiv.org/abs/2208.08243}{{\ttfamily 2208.08243}}].

\bibitem{Palla2023}
M.~{Palla}, I.~{De Looze}, M.~{Rela{\~n}o}, S.~{van der Giessen}, P.~{Dayal}, A.~{Ferrara} et~al., \emph{{Metal and dust evolution in ALMA REBELS galaxies: insights for future JWST observations}}, \href{https://doi.org/10.48550/arXiv.2311.16071}{\emph{arXiv e-prints} (2023) arXiv:2311.16071} [\href{https://arxiv.org/abs/2311.16071}{{\ttfamily 2311.16071}}].

\bibitem{Bethermin2020}
M.~{B{\'e}thermin}, Y.~{Fudamoto}, M.~{Ginolfi}, F.~{Loiacono}, Y.~{Khusanova}, P.L.~{Capak} et~al., \emph{{The ALPINE-ALMA [CII] survey: Data processing, catalogs, and statistical source properties}}, \href{https://doi.org/10.1051/0004-6361/202037649}{\emph{A\&A} {\bfseries 643} (2020) A2} [\href{https://arxiv.org/abs/2002.00962}{{\ttfamily 2002.00962}}].

\bibitem{Faisst2020}
A.L.~{Faisst}, D.~{Schaerer}, B.C.~{Lemaux}, P.A.~{Oesch}, Y.~{Fudamoto}, P.~{Cassata} et~al., \emph{{The ALPINE-ALMA [C II] Survey: Multiwavelength Ancillary Data and Basic Physical Measurements}}, \href{https://doi.org/10.3847/1538-4365/ab7ccd}{\emph{ApJS,} {\bfseries 247} (2020) 61} [\href{https://arxiv.org/abs/1912.01621}{{\ttfamily 1912.01621}}].

\bibitem{Mitsuhashi2023b}
I.~{Mitsuhashi}, K.-i.~{Tadaki}, R.~{Ikeda}, R.~{Herrera-Camus}, M.~{Aravena}, I.~{De Looze} et~al., \emph{{The ALMA-CRISTAL survey: Widespread dust-obscured star formation in typical star-forming galaxies at z=4-6}}, \href{https://doi.org/10.48550/arXiv.2311.17671}{\emph{arXiv e-prints} (2023) arXiv:2311.17671} [\href{https://arxiv.org/abs/2311.17671}{{\ttfamily 2311.17671}}].

\bibitem{Mitsuhashi2023}
I.~{Mitsuhashi}, Y.~{Harikane}, F.E.~{Bauer}, T.~{Bakx}, A.~{Ferrara}, S.~{Fujimoto} et~al., \emph{{SERENADE II: An ALMA Multi-Band Dust-Continuum Analysis of 28 Galaxies at $5<z<8$ and the Physical Origin of the Dust Temperature Evolution}}, \href{https://doi.org/10.48550/arXiv.2311.16857}{\emph{arXiv e-prints} (2023) arXiv:2311.16857} [\href{https://arxiv.org/abs/2311.16857}{{\ttfamily 2311.16857}}].

\bibitem{Bouwens2022}
R.J.~{Bouwens}, R.~{Smit}, S.~{Schouws}, M.~{Stefanon}, R.~{Bowler}, R.~{Endsley} et~al., \emph{{Reionization Era Bright Emission Line Survey: Selection and Characterization of Luminous Interstellar Medium Reservoirs in the z > 6.5 Universe}}, \href{https://doi.org/10.3847/1538-4357/ac5a4a}{\emph{ApJ,} {\bfseries 931} (2022) 160} [\href{https://arxiv.org/abs/2106.13719}{{\ttfamily 2106.13719}}].

\bibitem{Nozawa2003}
T.~{Nozawa}, T.~{Kozasa}, H.~{Umeda}, K.~{Maeda} and K.~{Nomoto}, \emph{{Dust in the Early Universe: Dust Formation in the Ejecta of Population III Supernovae}}, \href{https://doi.org/10.1086/379011}{\emph{ApJ,} {\bfseries 598} (2003) 785} [\href{https://arxiv.org/abs/astro-ph/0307108}{{\ttfamily astro-ph/0307108}}].

\bibitem{Brooker2022}
E.S.~{Brooker}, S.M.~{Stangl}, C.M.~{Mauney} and C.L.~{Fryer}, \emph{{Dependence of Dust Formation on the Supernova Explosion}}, \href{https://doi.org/10.3847/1538-4357/ac57c3}{\emph{ApJ,} {\bfseries 931} (2022) 85} [\href{https://arxiv.org/abs/2103.12781}{{\ttfamily 2103.12781}}].

\bibitem{Valiante2009}
R.~{Valiante}, R.~{Schneider}, S.~{Bianchi} and A.C.~{Andersen}, \emph{{Stellar sources of dust in the high-redshift Universe}}, \href{https://doi.org/10.1111/j.1365-2966.2009.15076.x}{\emph{MNRAS,} {\bfseries 397} (2009) 1661} [\href{https://arxiv.org/abs/0905.1691}{{\ttfamily 0905.1691}}].

\bibitem{Mancini2015}
M.~{Mancini}, R.~{Schneider}, L.~{Graziani}, R.~{Valiante}, P.~{Dayal}, U.~{Maio} et~al., \emph{{The dust mass in z > 6 normal star-forming galaxies.}}, \href{https://doi.org/10.1093/MNRAS,l/slv070}{\emph{MNRAS,} {\bfseries 451} (2015) L70} [\href{https://arxiv.org/abs/1505.01841}{{\ttfamily 1505.01841}}].

\bibitem{Angle-Alcazar2017}
D.~{Angl{\'e}s-Alc{\'a}zar}, C.-A.~{Faucher-Gigu{\`e}re}, D.~{Kere{\v{s}}}, P.F.~{Hopkins}, E.~{Quataert} and N.~{Murray}, \emph{{The cosmic baryon cycle and galaxy mass assembly in the FIRE simulations}}, \href{https://doi.org/10.1093/MNRAS,/stx1517}{\emph{MNRAS,} {\bfseries 470} (2017) 4698} [\href{https://arxiv.org/abs/1610.08523}{{\ttfamily 1610.08523}}].

\bibitem{Draine2003}
B.T.~{Draine}, \emph{{Interstellar Dust Grains}}, \href{https://doi.org/10.1146/annurev.astro.41.011802.094840}{\emph{ARA\&A} {\bfseries 41} (2003) 241} [\href{https://arxiv.org/abs/astro-ph/0304489}{{\ttfamily astro-ph/0304489}}].

\bibitem{Narayanan2018}
D.~{Narayanan}, R.~{Dav{\'e}}, B.D.~{Johnson}, R.~{Thompson}, C.~{Conroy} and J.~{Geach}, \emph{{The IRX-{\ensuremath{\beta}} dust attenuation relation in cosmological galaxy formation simulations}}, \href{https://doi.org/10.1093/MNRAS,/stx2860}{\emph{MNRAS,} {\bfseries 474} (2018) 1718} [\href{https://arxiv.org/abs/1705.05858}{{\ttfamily 1705.05858}}].

\bibitem{Vogelsberger2020}
M.~{Vogelsberger}, D.~{Nelson}, A.~{Pillepich}, X.~{Shen}, F.~{Marinacci}, V.~{Springel} et~al., \emph{{High-redshift JWST predictions from IllustrisTNG: dust modelling and galaxy luminosity functions}}, \href{https://doi.org/10.1093/MNRAS,/staa137}{\emph{MNRAS,} {\bfseries 492} (2020) 5167} [\href{https://arxiv.org/abs/1904.07238}{{\ttfamily 1904.07238}}].

\bibitem{Mushtaq2023}
M.~{Mushtaq}, D.~{Ceverino}, R.S.~{Klessen}, S.~{Reissl} and P.H.~{Puttasiddappa}, \emph{{Dust attenuation in galaxies at cosmic dawn from the FirstLight simulations}}, \href{https://doi.org/10.48550/arXiv.2304.10150}{\emph{arXiv e-prints} (2023) arXiv:2304.10150} [\href{https://arxiv.org/abs/2304.10150}{{\ttfamily 2304.10150}}].

\bibitem{Narayanan2023}
D.~{Narayanan}, J.D.T.~{Smith}, B.S.~{Hensley}, Q.~{Li}, C.-Y.~{Hu}, K.~{Sandstrom} et~al., \emph{{A Framework for Modeling Polycyclic Aromatic Hydrocarbon Emission in Galaxy Evolution Simulations}}, \href{https://doi.org/10.3847/1538-4357/accf8d}{\emph{ApJ,} {\bfseries 951} (2023) 100} [\href{https://arxiv.org/abs/2301.07136}{{\ttfamily 2301.07136}}].

\bibitem{dacunha2013}
E.~{da Cunha}, B.~{Groves}, F.~{Walter}, R.~{Decarli}, A.~{Weiss}, F.~{Bertoldi} et~al., \emph{{On the Effect of the Cosmic Microwave Background in High-redshift (Sub-)millimeter Observations}}, \href{https://doi.org/10.1088/0004-637X/766/1/13}{\emph{ApJ,} {\bfseries 766} (2013) 13} [\href{https://arxiv.org/abs/1302.0844}{{\ttfamily 1302.0844}}].

\bibitem{Imara2018}
N.~{Imara}, A.~{Loeb}, B.D.~{Johnson}, C.~{Conroy} and P.~{Behroozi}, \emph{{A Model Connecting Galaxy Masses, Star Formation Rates, and Dust Temperatures across Cosmic Time}}, \href{https://doi.org/10.3847/1538-4357/aaa3f0}{\emph{ApJ,} {\bfseries 854} (2018) 36} [\href{https://arxiv.org/abs/1801.01499}{{\ttfamily 1801.01499}}].

\bibitem{Inoue2020}
A.K.~{Inoue}, T.~{Hashimoto}, H.~{Chihara} and C.~{Koike}, \emph{{Radiative equilibrium estimates of dust temperature and mass in high-redshift galaxies}}, \href{https://doi.org/10.1093/MNRAS,/staa1203}{\emph{MNRAS,} {\bfseries 495} (2020) 1577} [\href{https://arxiv.org/abs/2004.12612}{{\ttfamily 2004.12612}}].

\bibitem{Mauerhofer2023}
V.~{Mauerhofer} and P.~{Dayal}, \emph{{The dust enrichment and detectability of early galaxies in the JWST and ALMA era}}, \href{https://doi.org/10.48550/arXiv.2305.01681}{\emph{arXiv e-prints} (2023) arXiv:2305.01681} [\href{https://arxiv.org/abs/2305.01681}{{\ttfamily 2305.01681}}].

\bibitem{Calura2008}
F.~{Calura}, A.~{Pipino} and F.~{Matteucci}, \emph{{The cycle of interstellar dust in galaxies of different morphological types}}, \href{https://doi.org/10.1051/0004-6361:20078090}{\emph{A\&A} {\bfseries 479} (2008) 669} [\href{https://arxiv.org/abs/0706.2197}{{\ttfamily 0706.2197}}].

\bibitem{Mattsson2022}
L.~{Mattsson} and R.~{Hedvall}, \emph{{Acceleration and clustering of cosmic dust in a gravoturbulent gas I. Numerical simulation of the nearly Jeans-unstable case}}, \href{https://doi.org/10.1093/mnras/stab3216}{\emph{MNRAS} {\bfseries 509} (2022) 3660} [\href{https://arxiv.org/abs/2111.01289}{{\ttfamily 2111.01289}}].

\bibitem{McKinnon2017}
R.~{McKinnon}, P.~{Torrey}, M.~{Vogelsberger}, C.C.~{Hayward} and F.~{Marinacci}, \emph{{Simulating the dust content of galaxies: successes and failures}}, \href{https://doi.org/10.1093/mnras/stx467}{\emph{MNRAS} {\bfseries 468} (2017) 1505} [\href{https://arxiv.org/abs/1606.02714}{{\ttfamily 1606.02714}}].

\bibitem{Hou2019}
K.-C.~{Hou}, S.~{Aoyama}, H.~{Hirashita}, K.~{Nagamine} and I.~{Shimizu}, \emph{{Dust scaling relations in a cosmological simulation}}, \href{https://doi.org/10.1093/mnras/stz121}{\emph{MNRAS} {\bfseries 485} (2019) 1727} [\href{https://arxiv.org/abs/1901.02886}{{\ttfamily 1901.02886}}].

\bibitem{Parente2022}
M.~{Parente}, C.~{Ragone-Figueroa}, G.L.~{Granato}, S.~{Borgani}, G.~{Murante}, M.~{Valentini} et~al., \emph{{Dust evolution with MUPPI in cosmological volumes}}, \href{https://doi.org/10.1093/mnras/stac1913}{\emph{MNRAS} {\bfseries 515} (2022) 2053} [\href{https://arxiv.org/abs/2204.11884}{{\ttfamily 2204.11884}}].

\bibitem{Furlanetto2017}
S.R.~{Furlanetto}, J.~{Mirocha}, R.H.~{Mebane} and G.~{Sun}, \emph{{A minimalist feedback-regulated model for galaxy formation during the epoch of reionization}}, \href{https://doi.org/10.1093/MNRAS,/stx2132}{\emph{MNRAS,} {\bfseries 472} (2017) 1576} [\href{https://arxiv.org/abs/1611.01169}{{\ttfamily 1611.01169}}].

\bibitem{Trenti2010}
M.~{Trenti}, M.~{Stiavelli}, R.J.~{Bouwens}, P.~{Oesch}, J.M.~{Shull}, G.D.~{Illingworth} et~al., \emph{{The Galaxy Luminosity Function During the Reionization Epoch}}, \href{https://doi.org/10.1088/2041-8205/714/2/L202}{\emph{ApJL} {\bfseries 714} (2010) L202} [\href{https://arxiv.org/abs/1004.0384}{{\ttfamily 1004.0384}}].

\bibitem{Dayal2014}
P.~{Dayal}, A.~{Ferrara}, J.S.~{Dunlop} and F.~{Pacucci}, \emph{{Essential physics of early galaxy formation}}, \href{https://doi.org/10.1093/mnras/stu1848}{\emph{MNRAS} {\bfseries 445} (2014) 2545} [\href{https://arxiv.org/abs/1405.4862}{{\ttfamily 1405.4862}}].

\bibitem{Mason2015}
C.A.~{Mason}, M.~{Trenti} and T.~{Treu}, \emph{{The Galaxy UV Luminosity Function before the Epoch of Reionization}}, \href{https://doi.org/10.1088/0004-637X/813/1/21}{\emph{ApJ} {\bfseries 813} (2015) 21} [\href{https://arxiv.org/abs/1508.01204}{{\ttfamily 1508.01204}}].

\bibitem{Planck2018}
{Planck Collaboration}, N.~{Aghanim}, Y.~{Akrami}, M.~{Ashdown}, J.~{Aumont}, C.~{Baccigalupi} et~al., \emph{{Planck 2018 results. VI. Cosmological parameters}}, \href{https://doi.org/10.1051/0004-6361/201833910}{\emph{A\&A} {\bfseries 641} (2020) A6} [\href{https://arxiv.org/abs/1807.06209}{{\ttfamily 1807.06209}}].

\bibitem{Oke1983}
J.B.~{Oke} and J.E.~{Gunn}, \emph{{Secondary standard stars for absolute spectrophotometry.}}, \href{https://doi.org/10.1086/160817}{\emph{ApJ,} {\bfseries 266} (1983) 713}.

\bibitem{Hegde&Wyatt2024}
S.~{Hegde}, M.M.~{Wyatt} and S.R.~{Furlanetto}, \emph{{A hidden population of active galactic nuclei can explain the overabundance of luminous $z>10$ objects observed by JWST}}, \href{https://doi.org/10.48550/arXiv.2405.01629}{\emph{arXiv e-prints} (2024) arXiv:2405.01629} [\href{https://arxiv.org/abs/2405.01629}{{\ttfamily 2405.01629}}].

\bibitem{Trac2015}
H.~{Trac}, R.~{Cen} and P.~{Mansfield}, \emph{{SCORCH I: The Galaxy-Halo Connection in the First Billion Years}}, \href{https://doi.org/10.1088/0004-637X/813/1/54}{\emph{ApJ,} {\bfseries 813} (2015) 54} [\href{https://arxiv.org/abs/1507.02685}{{\ttfamily 1507.02685}}].

\bibitem{Vale2004}
A.~{Vale} and J.P.~{Ostriker}, \emph{{Linking halo mass to galaxy luminosity}}, \href{https://doi.org/10.1111/j.1365-2966.2004.08059.x}{\emph{MNRAS,} {\bfseries 353} (2004) 189} [\href{https://arxiv.org/abs/astro-ph/0402500}{{\ttfamily astro-ph/0402500}}].

\bibitem{Goerdt2015}
T.~{Goerdt}, D.~{Ceverino}, A.~{Dekel} and R.~{Teyssier}, \emph{{Distribution of streaming rates into high-redshift galaxies}}, \href{https://doi.org/10.1093/mnras/stv2005}{\emph{MNRAS} {\bfseries 454} (2015) 637} [\href{https://arxiv.org/abs/1505.01486}{{\ttfamily 1505.01486}}].

\bibitem{Mirocha2021}
J.~{Mirocha}, P.~{La Plante} and A.~{Liu}, \emph{{The importance of galaxy formation histories in models of reionization}}, \href{https://doi.org/10.1093/mnras/stab1871}{\emph{MNRAS} {\bfseries 507} (2021) 3872} [\href{https://arxiv.org/abs/2012.09189}{{\ttfamily 2012.09189}}].

\bibitem{Keres2005}
D.~{Kere{\v{s}}}, N.~{Katz}, D.H.~{Weinberg} and R.~{Dav{\'e}}, \emph{{How do galaxies get their gas?}}, \href{https://doi.org/10.1111/j.1365-2966.2005.09451.x}{\emph{MNRAS} {\bfseries 363} (2005) 2} [\href{https://arxiv.org/abs/astro-ph/0407095}{{\ttfamily astro-ph/0407095}}].

\bibitem{Birnboim2007}
Y.~{Birnboim}, A.~{Dekel} and E.~{Neistein}, \emph{{Bursting and quenching in massive galaxies without major mergers or AGNs}}, \href{https://doi.org/10.1111/j.1365-2966.2007.12074.x}{\emph{MNRAS} {\bfseries 380} (2007) 339} [\href{https://arxiv.org/abs/astro-ph/0703435}{{\ttfamily astro-ph/0703435}}].

\bibitem{Faucher-Giguere2011}
C.-A.~{Faucher-Gigu{\`e}re}, D.~{Kere{\v{s}}} and C.-P.~{Ma}, \emph{{The baryonic assembly of dark matter haloes}}, \href{https://doi.org/10.1111/j.1365-2966.2011.19457.x}{\emph{MNRAS,} {\bfseries 417} (2011) 2982} [\href{https://arxiv.org/abs/1103.0001}{{\ttfamily 1103.0001}}].

\bibitem{Peng2010}
Y.-j.~{Peng}, S.J.~{Lilly}, K.~{Kova{\v{c}}}, M.~{Bolzonella}, L.~{Pozzetti}, A.~{Renzini} et~al., \emph{{Mass and Environment as Drivers of Galaxy Evolution in SDSS and zCOSMOS and the Origin of the Schechter Function}}, \href{https://doi.org/10.1088/0004-637X/721/1/193}{\emph{ApJ} {\bfseries 721} (2010) 193} [\href{https://arxiv.org/abs/1003.4747}{{\ttfamily 1003.4747}}].

\bibitem{Woo2013}
J.~{Woo}, A.~{Dekel}, S.M.~{Faber}, K.~{Noeske}, D.C.~{Koo}, B.F.~{Gerke} et~al., \emph{{Dependence of galaxy quenching on halo mass and distance from its centre}}, \href{https://doi.org/10.1093/mnras/sts274}{\emph{MNRAS} {\bfseries 428} (2013) 3306} [\href{https://arxiv.org/abs/1203.1625}{{\ttfamily 1203.1625}}].

\bibitem{Dave2012}
R.~{Dav{\'e}}, K.~{Finlator} and B.D.~{Oppenheimer}, \emph{{An analytic model for the evolution of the stellar, gas and metal content of galaxies}}, \href{https://doi.org/10.1111/j.1365-2966.2011.20148.x}{\emph{MNRAS} {\bfseries 421} (2012) 98} [\href{https://arxiv.org/abs/1108.0426}{{\ttfamily 1108.0426}}].

\bibitem{Larson1974mar}
R.B.~{Larson}, \emph{{Dynamical models for the formation and evolution of spherical galaxies}}, \href{https://doi.org/10.1093/mnras/166.3.585}{\emph{MNRAS} {\bfseries 166} (1974) 585}.

\bibitem{Larson1974nov}
R.B.~{Larson}, \emph{{Effects of supernovae on the early evolution of galaxies}}, \href{https://doi.org/10.1093/mnras/169.2.229}{\emph{MNRAS} {\bfseries 169} (1974) 229}.

\bibitem{Thompson2005}
T.A.~{Thompson}, E.~{Quataert} and N.~{Murray}, \emph{{Radiation Pressure-supported Starburst Disks and Active Galactic Nucleus Fueling}}, \href{https://doi.org/10.1086/431923}{\emph{ApJ} {\bfseries 630} (2005) 167} [\href{https://arxiv.org/abs/astro-ph/0503027}{{\ttfamily astro-ph/0503027}}].

\bibitem{Krumholz2018}
M.R.~{Krumholz}, B.~{Burkhart}, J.C.~{Forbes} and R.M.~{Crocker}, \emph{{A unified model for galactic discs: star formation, turbulence driving, and mass transport}}, \href{https://doi.org/10.1093/mnras/sty852}{\emph{MNRAS} {\bfseries 477} (2018) 2716} [\href{https://arxiv.org/abs/1706.00106}{{\ttfamily 1706.00106}}].

\bibitem{Madau2014}
P.~{Madau} and M.~{Dickinson}, \emph{{Cosmic Star-Formation History}}, \href{https://doi.org/10.1146/annurev-astro-081811-125615}{\emph{ARA\&A} {\bfseries 52} (2014) 415} [\href{https://arxiv.org/abs/1403.0007}{{\ttfamily 1403.0007}}].

\bibitem{Calura2017}
F.~{Calura}, F.~{Pozzi}, G.~{Cresci}, P.~{Santini}, C.~{Gruppioni}, L.~{Pozzetti} et~al., \emph{{The dust-to-stellar mass ratio as a valuable tool to probe the evolution of local and distant star-forming galaxies}}, \href{https://doi.org/10.1093/mnras/stw2749}{\emph{MNRAS} {\bfseries 465} (2017) 54} [\href{https://arxiv.org/abs/1610.08979}{{\ttfamily 1610.08979}}].

\bibitem{Millan-Irigoyen2020}


\bibitem{Todini2001}
P.~{Todini} and A.~{Ferrara}, \emph{{Dust formation in primordial Type II supernovae}}, \href{https://doi.org/10.1046/j.1365-8711.2001.04486.x}{\emph{MNRAS,} {\bfseries 325} (2001) 726} [\href{https://arxiv.org/abs/astro-ph/0009176}{{\ttfamily astro-ph/0009176}}].

\bibitem{Bianchi2007}
S.~{Bianchi} and R.~{Schneider}, \emph{{Dust formation and survival in supernova ejecta}}, \href{https://doi.org/10.1111/j.1365-2966.2007.11829.x}{\emph{MNRAS} {\bfseries 378} (2007) 973} [\href{https://arxiv.org/abs/0704.0586}{{\ttfamily 0704.0586}}].

\bibitem{Temim2017}
T.~{Temim}, E.~{Dwek}, R.G.~{Arendt}, K.J.~{Borkowski}, S.P.~{Reynolds}, P.~{Slane} et~al., \emph{{A Massive Shell of Supernova-formed Dust in SNR G54.1+0.3}}, \href{https://doi.org/10.3847/1538-4357/836/1/129}{\emph{ApJ} {\bfseries 836} (2017) 129} [\href{https://arxiv.org/abs/1701.01117}{{\ttfamily 1701.01117}}].

\bibitem{Rho2018}
J.~{Rho}, H.L.~{Gomez}, A.~{Boogert}, M.W.L.~{Smith}, P.O.~{Lagage}, D.~{Dowell} et~al., \emph{{A dust twin of Cas A: cool dust and 21 {\ensuremath{\mu}}m silicate dust feature in the supernova remnant G54.1+0.3}}, \href{https://doi.org/10.1093/mnras/sty1713}{\emph{MNRAS} {\bfseries 479} (2018) 5101} [\href{https://arxiv.org/abs/1707.08230}{{\ttfamily 1707.08230}}].

\bibitem{Burgarella2020}
D.~{Burgarella}, A.~{Nanni}, H.~{Hirashita}, P.~{Theul{\'e}}, A.K.~{Inoue} and T.T.~{Takeuchi}, \emph{{Observational and theoretical constraints on the formation and early evolution of the first dust grains in galaxies at 5 < z < 10}}, \href{https://doi.org/10.1051/0004-6361/201937143}{\emph{A\&A} {\bfseries 637} (2020) A32} [\href{https://arxiv.org/abs/2002.01858}{{\ttfamily 2002.01858}}].

\bibitem{Donevski2020}
D.~{Donevski}, A.~{Lapi}, K.~{Ma{\l}ek}, D.~{Liu}, C.~{G{\'o}mez-Guijarro}, R.~{Dav{\'e}} et~al., \emph{{In pursuit of giants. I. The evolution of the dust-to-stellar mass ratio in distant dusty galaxies}}, \href{https://doi.org/10.1051/0004-6361/202038405}{\emph{A\&A} {\bfseries 644} (2020) A144} [\href{https://arxiv.org/abs/2008.09995}{{\ttfamily 2008.09995}}].

\bibitem{Witstok2023}
J.~{Witstok}, G.C.~{Jones}, R.~{Maiolino}, R.~{Smit} and R.~{Schneider}, \emph{{An empirical study of dust properties at the earliest epochs}}, \href{https://doi.org/10.1093/mnras/stad1470}{\emph{MNRAS} {\bfseries 523} (2023) 3119} [\href{https://arxiv.org/abs/2305.09714}{{\ttfamily 2305.09714}}].

\bibitem{Cherchneff2010}
I.~{Cherchneff} and E.~{Dwek}, \emph{{The Chemistry of Population III Supernova Ejecta. II. The Nucleation of Molecular Clusters as a Diagnostic for Dust in the Early Universe}}, \href{https://doi.org/10.1088/0004-637X/713/1/1}{\emph{ApJ,} {\bfseries 713} (2010) 1} [\href{https://arxiv.org/abs/1002.3060}{{\ttfamily 1002.3060}}].

\bibitem{Furlanetto2022}
S.R.~{Furlanetto} and J.~{Mirocha}, \emph{{Bursty star formation during the Cosmic Dawn driven by delayed stellar feedback}}, \href{https://doi.org/10.1093/MNRAS,/stac310}{\emph{MNRAS,} {\bfseries 511} (2022) 3895} [\href{https://arxiv.org/abs/2109.04488}{{\ttfamily 2109.04488}}].

\bibitem{Matsuura2013}
M.~{Matsuura}, P.M.~{Woods} and P.J.~{Owen}, \emph{{The global gas and dust budget of the Small Magellanic Cloud}}, \href{https://doi.org/10.1093/MNRAS,/sts521}{\emph{MNRAS,} {\bfseries 429} (2013) 2527} [\href{https://arxiv.org/abs/1212.1468}{{\ttfamily 1212.1468}}].

\bibitem{Mirocha2020}
J.~{Mirocha}, C.~{Mason} and D.P.~{Stark}, \emph{{Effects of self-consistent rest-ultraviolet colours in semi-empirical galaxy formation models}}, \href{https://doi.org/10.1093/MNRAS,/staa2586}{\emph{MNRAS,} {\bfseries 498} (2020) 2645} [\href{https://arxiv.org/abs/2005.07208}{{\ttfamily 2005.07208}}].

\bibitem{Bouwens2016}
R.J.~{Bouwens}, M.~{Aravena}, R.~{Decarli}, F.~{Walter}, E.~{da Cunha}, I.~{Labb{\'e}} et~al., \emph{{ALMA Spectroscopic Survey in the Hubble Ultra Deep Field: The Infrared Excess of UV-Selected z = 2-10 Galaxies as a Function of UV-Continuum Slope and Stellar Mass}}, \href{https://doi.org/10.3847/1538-4357/833/1/72}{\emph{ApJ,} {\bfseries 833} (2016) 72} [\href{https://arxiv.org/abs/1606.05280}{{\ttfamily 1606.05280}}].

\bibitem{LoebFurlanetto2013}
A.~{Loeb} and S.R.~{Furlanetto}, \emph{{The First Galaxies in the Universe}} (2013).

\bibitem{Faucher-Giguere2018}
C.-A.~{Faucher-Gigu{\`e}re}, \emph{{A model for the origin of bursty star formation in galaxies}}, \href{https://doi.org/10.1093/MNRAS,/stx2595}{\emph{MNRAS,} {\bfseries 473} (2018) 3717} [\href{https://arxiv.org/abs/1701.04824}{{\ttfamily 1701.04824}}].

\bibitem{Pozzi2024}
F.~{Pozzi}, F.~{Calura}, Q.~{D'Amato}, M.~{Gavarente}, M.~{Bethermin}, M.~{Boquien} et~al., \emph{{The ALPINE-ALMA [CII] survey: Dust emission effective radius up to 3 kpc in the early Universe}}, \href{https://doi.org/10.1051/0004-6361/202348996}{\emph{A\&A} {\bfseries 686} (2024) A187} [\href{https://arxiv.org/abs/2403.13490}{{\ttfamily 2403.13490}}].

\bibitem{Varosi1999}
F.~{V{\'a}rosi} and E.~{Dwek}, \emph{{Analytical Approximations for Calculating the Escape and Absorption of Radiation in Clumpy Dusty Environments}}, \href{https://doi.org/10.1086/307729}{\emph{ApJ,} {\bfseries 523} (1999) 265} [\href{https://arxiv.org/abs/astro-ph/9905029}{{\ttfamily astro-ph/9905029}}].

\bibitem{Bianchi2019}
S.~{Bianchi}, V.~{Casasola}, M.~{Baes}, C.J.R.~{Clark}, E.~{Corbelli}, J.I.~{Davies} et~al., \emph{{Dust emissivity and absorption cross section in DustPedia late-type galaxies}}, \href{https://doi.org/10.1051/0004-6361/201936314}{\emph{A\&A} {\bfseries 631} (2019) A102} [\href{https://arxiv.org/abs/1909.12692}{{\ttfamily 1909.12692}}].

\bibitem{Bouwens2014}
R.J.~{Bouwens}, G.D.~{Illingworth}, P.A.~{Oesch}, I.~{Labb{\'e}}, P.G.~{van Dokkum}, M.~{Trenti} et~al., \emph{{UV-continuum Slopes of >4000 z \raisebox{-0.5ex}\textasciitilde 4-8 Galaxies from the HUDF/XDF, HUDF09, ERS, CANDELS-South, and CANDELS-North Fields}}, \href{https://doi.org/10.1088/0004-637X/793/2/115}{\emph{ApJ,} {\bfseries 793} (2014) 115} [\href{https://arxiv.org/abs/1306.2950}{{\ttfamily 1306.2950}}].

\bibitem{Cullen2023a}
F.~{Cullen}, D.J.~{McLeod}, R.J.~{McLure}, J.S.~{Dunlop}, C.T.~{Donnan}, A.C.~{Carnall} et~al., \emph{{Evidence for the emergence of dust-free stellar populations at z > 10}}, \href{https://doi.org/10.48550/arXiv.2311.06209}{\emph{arXiv e-prints} (2023) arXiv:2311.06209} [\href{https://arxiv.org/abs/2311.06209}{{\ttfamily 2311.06209}}].

\bibitem{Naidu2022}
R.P.~{Naidu}, P.A.~{Oesch}, D.J.~{Setton}, J.~{Matthee}, C.~{Conroy}, B.D.~{Johnson} et~al., \emph{{Schrodinger's Galaxy Candidate: Puzzlingly Luminous at $z\approx17$, or Dusty/Quenched at $z\approx5$?}}, \href{https://doi.org/10.48550/arXiv.2208.02794}{\emph{arXiv e-prints} (2022) arXiv:2208.02794} [\href{https://arxiv.org/abs/2208.02794}{{\ttfamily 2208.02794}}].

\bibitem{Furlanetto2023}
S.R.~{Furlanetto} and J.~{Mirocha}, \emph{{On the expected purity of photometric galaxy surveys targeting the Cosmic Dawn}}, \href{https://doi.org/10.1093/MNRAS,/stad1799}{\emph{MNRAS,} {\bfseries 523} (2023) 5274} [\href{https://arxiv.org/abs/2208.12828}{{\ttfamily 2208.12828}}].

\bibitem{Zavala2023}
J.A.~{Zavala}, V.~{Buat}, C.M.~{Casey}, S.L.~{Finkelstein}, D.~{Burgarella}, M.B.~{Bagley} et~al., \emph{{Dusty Starbursts Masquerading as Ultra-high Redshift Galaxies in JWST CEERS Observations}}, \href{https://doi.org/10.3847/2041-8213/acacfe}{\emph{ApJ,l} {\bfseries 943} (2023) L9} [\href{https://arxiv.org/abs/2208.01816}{{\ttfamily 2208.01816}}].

\bibitem{Cullen2023}
F.~{Cullen}, R.J.~{McLure}, D.J.~{McLeod}, J.S.~{Dunlop}, C.T.~{Donnan}, A.C.~{Carnall} et~al., \emph{{The ultraviolet continuum slopes ({\ensuremath{\beta}}) of galaxies at z $\simeq$ 8-16 from JWST and ground-based near-infrared imaging}}, \href{https://doi.org/10.1093/MNRAS,/stad073}{\emph{MNRAS,} {\bfseries 520} (2023) 14} [\href{https://arxiv.org/abs/2208.04914}{{\ttfamily 2208.04914}}].

\bibitem{Munoz2023}
J.B.~{Mu{\~n}oz}, J.~{Mirocha}, S.~{Furlanetto} and N.~{Sabti}, \emph{{Breaking degeneracies in the first galaxies with clustering}}, \href{https://doi.org/10.1093/MNRAS,l/slad115}{\emph{MNRAS,} {\bfseries 526} (2023) L47} [\href{https://arxiv.org/abs/2306.09403}{{\ttfamily 2306.09403}}].

\bibitem{Bland-Hawthorn2003}
J.~{Bland-Hawthorn} and M.~{Cohen}, \emph{{The Large-Scale Bipolar Wind in the Galactic Center}}, \href{https://doi.org/10.1086/344573}{\emph{ApJ} {\bfseries 582} (2003) 246} [\href{https://arxiv.org/abs/astro-ph/0208553}{{\ttfamily astro-ph/0208553}}].

\bibitem{Rupke2013}
D.S.N.~{Rupke} and S.~{Veilleux}, \emph{{The Multiphase Structure and Power Sources of Galactic Winds in Major Mergers}}, \href{https://doi.org/10.1088/0004-637X/768/1/75}{\emph{ApJ} {\bfseries 768} (2013) 75} [\href{https://arxiv.org/abs/1303.6866}{{\ttfamily 1303.6866}}].

\bibitem{Triani2021}
D.P.~{Triani}, M.~{Sinha}, D.J.~{Croton}, E.~{Dwek} and C.~{Pacifici}, \emph{{Exploring the relation between dust mass and galaxy properties using Dusty SAGE}}, \href{https://doi.org/10.1093/mnras/stab558}{\emph{MNRAS} {\bfseries 503} (2021) 1005} [\href{https://arxiv.org/abs/2102.12652}{{\ttfamily 2102.12652}}].

\bibitem{Katsioli2023}
S.~{Katsioli}, E.M.~{Xilouris}, C.~{Kramer}, R.~{Adam}, P.~{Ade}, H.~{Ajeddig} et~al., \emph{{The stratification of ISM properties in the edge-on galaxy NGC 891 revealed by NIKA2}}, \href{https://doi.org/10.1051/0004-6361/202347020}{\emph{A\&A} {\bfseries 679} (2023) A7} [\href{https://arxiv.org/abs/2309.08403}{{\ttfamily 2309.08403}}].

\bibitem{Topping2022}
M.W.~{Topping}, D.P.~{Stark}, R.~{Endsley}, R.J.~{Bouwens}, S.~{Schouws}, R.~{Smit} et~al., \emph{{The ALMA REBELS Survey: specific star formation rates in the reionization era}}, \href{https://doi.org/10.1093/mnras/stac2291}{\emph{MNRAS} {\bfseries 516} (2022) 975} [\href{https://arxiv.org/abs/2203.07392}{{\ttfamily 2203.07392}}].

\bibitem{Sommovigo2022}
L.~{Sommovigo}, A.~{Ferrara}, A.~{Pallottini}, P.~{Dayal}, R.J.~{Bouwens}, R.~{Smit} et~al., \emph{{The ALMA REBELS Survey: cosmic dust temperature evolution out to z 7}}, \href{https://doi.org/10.1093/MNRAS,/stac302}{\emph{MNRAS,} {\bfseries 513} (2022) 3122} [\href{https://arxiv.org/abs/2202.01227}{{\ttfamily 2202.01227}}].

\bibitem{deRossi2023}
M.E.~{De Rossi} and V.~{Bromm}, \emph{{Probing the Nature of the First Galaxies with JWST and ALMA}}, \href{https://doi.org/10.3847/2041-8213/acc32e}{\emph{ApJL} {\bfseries 946} (2023) L20} [\href{https://arxiv.org/abs/2303.06328}{{\ttfamily 2303.06328}}].

\bibitem{Dekel2023}
A.~{Dekel}, K.C.~{Sarkar}, Y.~{Birnboim}, N.~{Mandelker} and Z.~{Li}, \emph{{Efficient formation of massive galaxies at cosmic dawn by feedback-free starbursts}}, \href{https://doi.org/10.1093/MNRAS,/stad1557}{\emph{MNRAS,} {\bfseries 523} (2023) 3201} [\href{https://arxiv.org/abs/2303.04827}{{\ttfamily 2303.04827}}].

\bibitem{Tsuna2023}
D.~{Tsuna}, Y.~{Nakazato} and T.~{Hartwig}, \emph{{A Photon Burst Clears the Earliest Dusty Galaxies: Modelling Dust in High-redshift Galaxies from ALMA to JWST}}, \href{https://doi.org/10.48550/arXiv.2309.02415}{\emph{arXiv e-prints} (2023) arXiv:2309.02415} [\href{https://arxiv.org/abs/2309.02415}{{\ttfamily 2309.02415}}].

\bibitem{Barkana2001}
R.~{Barkana} and A.~{Loeb}, \emph{{In the beginning: the first sources of light and the reionization of the universe}}, \href{https://doi.org/10.1016/S0370-1573(01)00019-9}{\emph{Physics Reports} {\bfseries 349} (2001) 125} [\href{https://arxiv.org/abs/astro-ph/0010468}{{\ttfamily astro-ph/0010468}}].

\bibitem{Simmonds2023}
C.~{Simmonds}, S.~{Tacchella}, K.~{Hainline}, B.D.~{Johnson}, W.~{McClymont}, B.~{Robertson} et~al., \emph{{Low-mass bursty galaxies in JADES efficiently produce ionising photons and could represent the main drivers of reionisation}}, \href{https://doi.org/10.48550/arXiv.2310.01112}{\emph{arXiv e-prints} (2023) arXiv:2310.01112} [\href{https://arxiv.org/abs/2310.01112}{{\ttfamily 2310.01112}}].

\bibitem{Williams2023}
C.C.~{Williams}, S.~{Alberts}, Z.~{Ji}, K.N.~{Hainline}, J.~{Lyu}, G.~{Rieke} et~al., \emph{{The galaxies missed by Hubble and ALMA: the contribution of extremely red galaxies to the cosmic census at 3<z<8}}, {\emph{arXiv e-prints} (2023) arXiv:2311.07483} [\href{https://arxiv.org/abs/2311.07483}{{\ttfamily 2311.07483}}].

\bibitem{Richie2024}
H.M.~{Richie}, E.E.~{Schneider}, M.W.~{Abruzzo} and P.~{Torrey}, \emph{{Dust Survival in Galactic Winds}}, \href{https://doi.org/10.48550/arXiv.2403.03711}{\emph{arXiv e-prints} (2024) arXiv:2403.03711} [\href{https://arxiv.org/abs/2403.03711}{{\ttfamily 2403.03711}}].

\bibitem{Yamaguchi2023}
N.~{Yamaguchi}, S.R.~{Furlanetto} and A.C.~{Trapp}, \emph{{The extent of intergalactic metal enrichment from galactic winds during the Cosmic Dawn}}, \href{https://doi.org/10.1093/MNRAS,/stad315}{\emph{MNRAS,} {\bfseries 520} (2023) 2922} [\href{https://arxiv.org/abs/2209.09345}{{\ttfamily 2209.09345}}].

\bibitem{Donnan2023}
C.T.~{Donnan}, D.J.~{McLeod}, J.S.~{Dunlop}, R.J.~{McLure}, A.C.~{Carnall}, R.~{Begley} et~al., \emph{{The evolution of the galaxy UV luminosity function at redshifts z $\simeq$ 8 - 15 from deep JWST and ground-based near-infrared imaging}}, \href{https://doi.org/10.1093/MNRAS,/stac3472}{\emph{MNRAS,} {\bfseries 518} (2023) 6011} [\href{https://arxiv.org/abs/2207.12356}{{\ttfamily 2207.12356}}].

\bibitem{Donnan2023a}
C.T.~{Donnan}, D.J.~{McLeod}, R.J.~{McLure}, J.S.~{Dunlop}, A.C.~{Carnall}, F.~{Cullen} et~al., \emph{{The abundance of z {\ensuremath{\gtrsim}} 10 galaxy candidates in the HUDF using deep JWST NIRCam medium-band imaging}}, \href{https://doi.org/10.1093/MNRAS,/stad471}{\emph{MNRAS,} {\bfseries 520} (2023) 4554} [\href{https://arxiv.org/abs/2212.10126}{{\ttfamily 2212.10126}}].

\bibitem{Matplotlib}
J.D.~Hunter, \emph{Matplotlib: A 2d graphics environment}, \href{https://doi.org/10.1109/MCSE.2007.55}{\emph{Computing in Science \& Engineering} {\bfseries 9} (2007) 90}.

\bibitem{Numpy}
C.R.~Harris, K.J.~Millman, S.J.~van~der Walt, R.~Gommers, P.~Virtanen, D.~Cournapeau et~al., \emph{Array programming with {NumPy}}, \href{https://doi.org/10.1038/s41586-020-2649-2}{\emph{Nature} {\bfseries 585} (2020) 357}.

\bibitem{Pandas}
T.~pandas~development team, \emph{pandas-dev/pandas: Pandas},  Feb., 2020.
\newblock 10.5281/zenodo.3509134.

\bibitem{Scipy}
P.~Virtanen, R.~Gommers, T.E.~Oliphant, M.~Haberland, T.~Reddy, D.~Cournapeau et~al., \emph{{{SciPy} 1.0: Fundamental Algorithms for Scientific Computing in Python}}, \href{https://doi.org/10.1038/s41592-019-0686-2}{\emph{Nature Methods} {\bfseries 17} (2020) 261}.

\end{thebibliography}\endgroup
\end{document}